\documentclass[%
 reprint,
 amsmath,
 amssymb,
 showkeys,
 pra,
 floatfix,
]{revtex4-2}

\usepackage{xr}

\usepackage{titlesec}

\usepackage{amsmath} 
\usepackage[linesnumbered,ruled,vlined]{algorithm2e}
\let\oldnl\nl
\newcommand{\nonl}{\renewcommand{\nl}{\let\nl\oldnl}}
\usepackage{braket}
\usepackage{float}
\usepackage[table,xcdraw]{xcolor}
\usepackage{graphicx}
\usepackage{multirow}
\usepackage{dcolumn}
\usepackage{enumitem}
\usepackage{lipsum}  
\usepackage{bm}
\usepackage{hyperref}
\hypersetup{
    colorlinks=true,
    linkcolor=blue,
    citecolor=magenta
}
\usepackage[T1]{fontenc}
\usepackage[latin1]{inputenc}
\usepackage[table]{xcolor}
\usepackage{etoolbox}
\usepackage{pgf} 
\usepackage{subcaption}
\usepackage{array}
\makeatletter
\newcommand\subsubsubsection{\@startsection{paragraph}{4}{\z@}{-2.5ex\@plus -1ex \@minus -.25ex}{1.25ex \@plus .25ex}{\normalfont\footnotesize\bfseries}}
\newcommand\subsubsubsubsection{\@startsection{subparagraph}{5}{\z@}{-2.5ex\@plus -1ex \@minus -.25ex}{1.25ex \@plus .25ex}{\normalfont\footnotesize\bfseries}}
\makeatother            

\begin{document}

\preprint{APS/123-QED}

\title{Quantum Phase Recognition using Quantum Tensor Networks}
\author{Shweta Sahoo}
\author{Utkarsh Azad}
\email{utkarsh.azad@research.iiit.ac.in}
\author{Harjinder Singh}

\affiliation{%
    Center for Computational Natural Sciences and Bioinformatics, International Institute of Information Technology, Hyderabad.\\
    Center for Quantum Science and Technology,\\ International Institute of Information Technology, Hyderabad.
}%

\date{\today}

\begin{abstract}
   Machine learning (ML) has recently facilitated many advances in solving problems related to many-body physical systems. Given the intrinsic quantum nature of these problems, it is natural to speculate that quantum-enhanced machine learning will enable us to unveil even greater details than we currently have. With this motivation, this paper examines a quantum machine learning approach based on shallow variational ansatz inspired by tensor networks for supervised learning tasks. In particular, we first look at the standard image classification tasks using the Fashion-MNIST dataset and study the effect of repeating tensor network layers on ansatz's expressibility and performance. Finally, we use this strategy to tackle the problem of quantum phase recognition for the transverse-field Ising and Heisenberg spin models in one and two dimensions, where we were able to reach $\geq 98\%$ test-set accuracies with both multi-scale entanglement renormalization ansatz (MERA) and tree tensor network (TTN) inspired parametrized quantum circuits. 
\end{abstract}

\maketitle

\begin{figure*}[t]
    \centering
    \begin{subfigure}[b]{0.46\linewidth}
    \hspace{-20pt}
    \begin{minipage}
    {.03\textwidth}
        \caption{}
        \label{fig:ttn-circuit}
    \end{minipage}%
    \begin{minipage}{0.90\textwidth}
        \includegraphics[width=.92\linewidth]{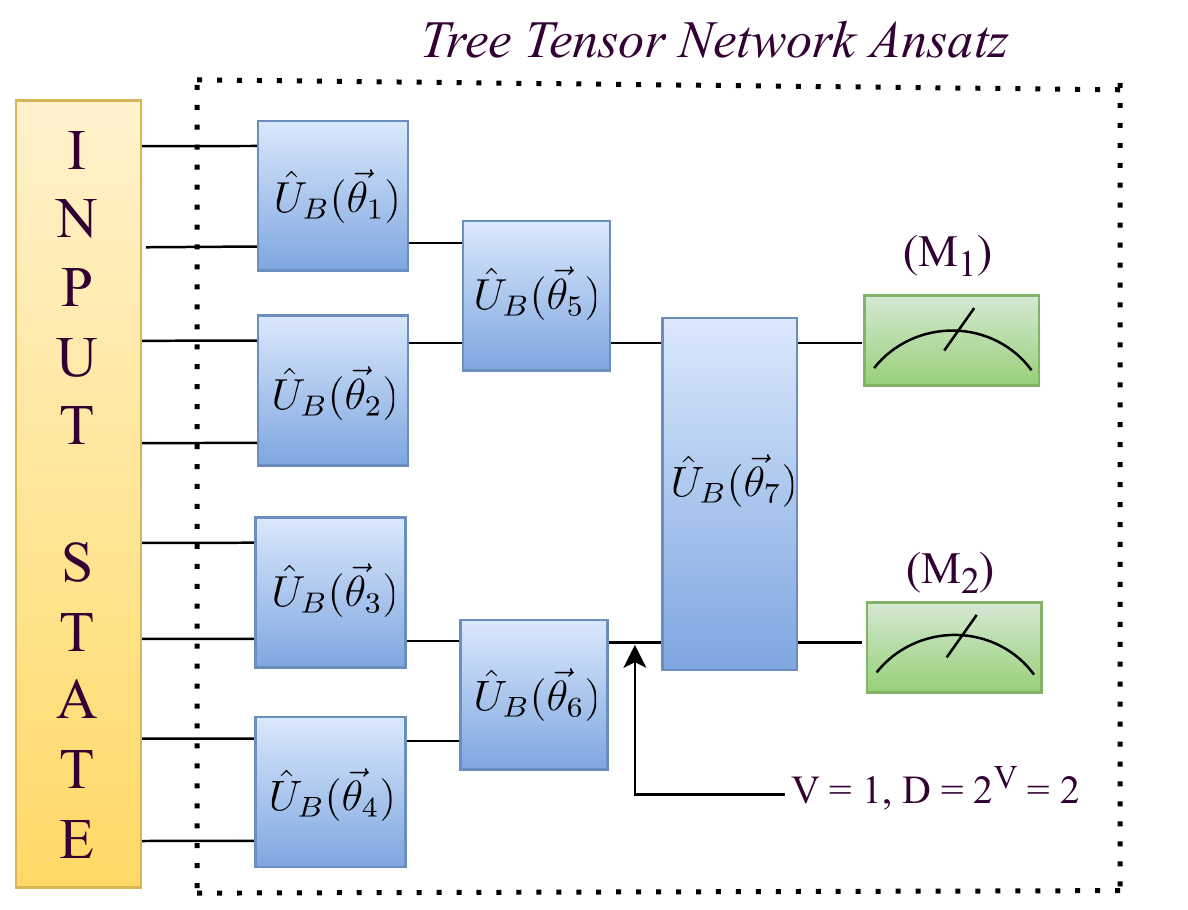}
    \end{minipage}
    \end{subfigure}
    \begin{subfigure}[b]{0.52\linewidth}
    \hspace{-35pt}
    \begin{minipage}{.08\textwidth}
        \caption{}
        \label{fig:mera-circuit}
    \end{minipage}%
    \begin{minipage}{0.98\textwidth}
        \includegraphics[width=\linewidth]{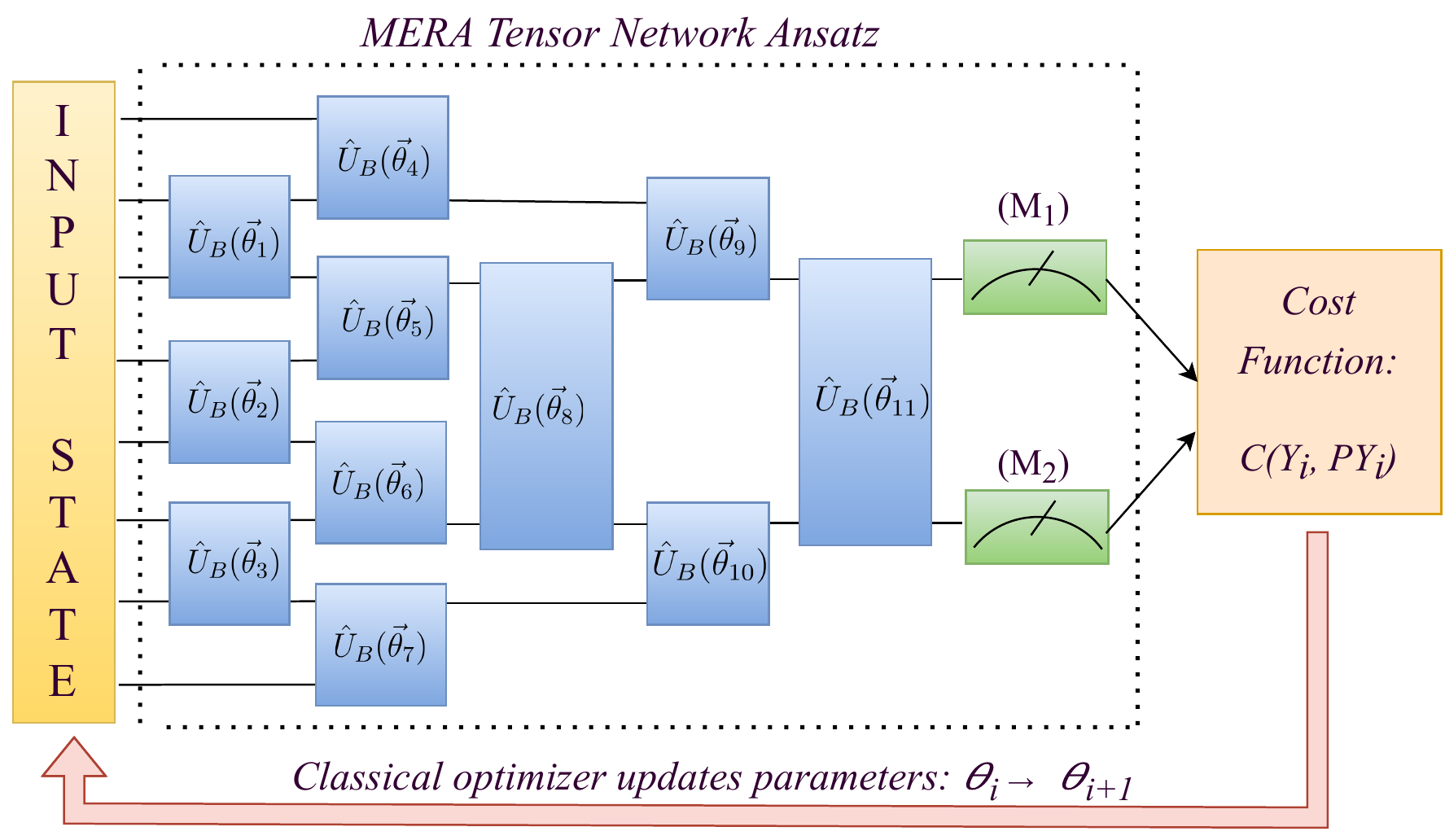}
    \end{minipage}
    \end{subfigure}
    \caption{\textbf{Tensor network inspired variational ans\"{a}tze}: (a) Structure of the tree tensor network (TTN) ansatz with bond dimension, $D=2$, and (b)  variational workflow using multi-scale entanglement renormalization ansatz (MERA) tensor network.} 
    \label{fig:vqe-circuit}
\end{figure*}



\section{\label{sec:intro} Introduction}

Machine learning (ML) offers tools and techniques to learn and predict patterns that emerge in data. One crucial avenue of this pattern recognition task is classification, which involves predicting class labels for the input data and has found applications in speech recognition \cite{Deng2013}, biometric identification \cite{https://doi.org/10.48550/arxiv.2104.03255}, object classification \cite{https://doi.org/10.48550/arxiv.1102.0183}, disease identification \cite{Amrane2018} and many more. These applications result from immense leaps classical ML algorithms have made in dealing with various challenging datasets. Lately, based on these successes, ML algorithms have been used for problems related to many-body physical systems, such as recognizing phases of matter \cite{Carrasquilla2017, vanNieuwenburg2017}. Even though they have shown more promising results for studying relevant and useful many-body systems than the best contemporary classical algorithms, they still do not alleviate the sign problem \cite{PhysRevB.41.9301}, which usually emerges in such calculations and causes an exponential slowdown.

More recently, there has been an ongoing effort to develop quantum-enhanced machine learning algorithms that leverage quantum computers to tackle traditional ML problems \cite{https://doi.org/10.48550/arxiv.1802.06002}. These algorithms are typically based on a class of hybrid quantum-classical algorithms called variational quantum algorithms (VQAs), such as variational quantum eigensolver (VQE) \cite{Peruzzo2014} and variational quantum linear solvers (VQLS) \cite{2019arXiv190905820B} that have been used to find the ground state of a Hamiltonian and solve systems of linear equations on noisy intermediate-scale quantum (NISQ) hardware \cite{Preskill2018}. The relatively short depth of the parameterized quantum circuits (PQCs) used in these algorithms makes them an ideal candidate for achieving good results on NISQ devices without error correction codes. \cite{2020arXiv201209265C}. 

In principle, although PQCs are analogous to classical neural networks structurally, they can exploit additional computational resources due to the presence of quantum mechanical phenomena such as superposition and entanglement \cite{Azad2022}. The basic working principle of VQAs is to optimize the parameters of PQC, also referred to as an \textit{ansatz}, using a classical optimization routine to minimize a cost function defined on measurements taken on the qubits present in the ansatz. Therefore, the performance of these algorithms is majorly based on the structure of the ansatz \cite{2022arXiv220502095A}. Hence, it is crucial to analyze and have some basic insights into the ansatz for a particular problem or application to assess and improve their trainability.  

In this paper, we use a VQE-based algorithm to classify classical and quantum data. For the former, we look at the task of classification of Fashion MNIST dataset \cite{https://doi.org/10.48550/arxiv.1708.07747}, whereas, for the latter, we tackle the problem of classification of the quantum phase of 1-D and 2-D transverse-field Ising and XXZ Heisenberg spin system. We employ multi-scale entanglement renormalization ansatz (MERA) and tree tensor network (TTN) states for building the ansatz for the variational routine. Finally, we also use expressibility and entangling capability analysis for choosing the structure of unitary block for these ans\"{a}tze (Figs. \ref{fig:su4-block-circuit} and \ref{fig:small-block-circuit}) that helps us make use of shorter-depth blocks than the general $SU(4)$ one suggested in \cite{Lazzarin2022}.

\textit{Structure} : In section \ref{sec:background}, we start off with a background on quantum tensor networks and spin systems, followed by a description of the experiments and their corresponding results conducted by us in section \ref{sec:setup}. Lastly, in section \ref{sec:conclusions}, we provide our conclusions and discussions on our results and future work.   

\section{\label{sec:background} Background}

\subsection{\label{sec:bg:qtn}Quantum Tensor Networks}
Tensor networks are essentially approximations of very large tensors using smaller, easier to handle tensors \cite{https://doi.org/10.48550/arxiv.1605.05775}. Tensor networks like the matrix product state (MPS) \cite{Ors2014}, tree tensor networks (TTN) \cite{Shi2006} and multi-scale entanglement renormalization ansatz (MERA) \cite{https://doi.org/10.48550/arxiv.quant-ph/0610099} can be constructed using a quantum circuit \cite{Huggins2019, Haghshenas2022}. 

Recently, the use of quantum circuits based on tensor networks have been explored in the domain of machine learning for both generative \cite{Wall2021} and discriminative tasks \cite{Huggins2019, Grant2018, Lazzarin2022}. The bond dimension of a tensor network is the dimension of the index connecting smaller tensors together. The bond dimension, $D$, of a tensor network that has been realized using a quantum circuit is $2^{v}$, where $v$ qubits connect different subtrees. The main motivation behind using tensor network inspired quantum circuits is the increase in expressibility of the ansatz with increasing bond dimension to the point that the entire state space can be covered for a sufficiently huge bond dimension. In the classical scenario, such systems will be too computationally expensive to deal with, as presented in \cite{Huggins2019}.
 Moreover, it is highly likely that in the case of quantum data, such as the wavefunction of a system, usage of classical methods will be intractable due to the exponential increase of information that needs to be encoded and computed with the increase in particles. 

Circuits with a hierarchical structure like that of a MERA or TTN tensor network have been used in the classification of images like those in the MNIST dataset \cite{Grant2018, Lazzarin2022}. A hybrid classical-quantum MPS-VQC has been used in image classification tasks with the MPS being the feature extractor for the images. While the MPS tensor network in this case is classical in nature, it can be replaced with an equivalent quantum circuit paving the way for the usage of quantum tensor networks as feature extractors \cite{2020arXiv201114651Y}.
Various tensor network ans\"{a}tze have also been used for quantum phase recognition tasks for the 1-D Heisenberg \cite{Lazzarin2022} and transverse-field Ising models \cite{PhysRevA.102.012415} with good results.

\subsection{\label{sec:bg:tl}Spin Systems}
The study of spin systems is important in order to understand the magnetic properties of a system at a macroscopic level. This is because the magnetic moment of an atom has contributions from the electron spins. The alignment of many such spins on a macroscopic scale defines the magnetic properties of the system. This alignment of spins is driven by the exchange interaction between the atoms. Exchange interaction is a short-range, powerful interaction that occurs due to the electrical forces between electrons in the atoms \cite{Parkinson2010, ohanyan}. In this work, we will deal with spin systems where the atomic dipoles are depicted by points on 1-D (chain) and 2-D (rectangular) lattices. The exchange interaction between atoms is limited to the nearest neighbors and is given by the general formula:

\[\varepsilon = \alpha\boldsymbol{S_1}{.}\boldsymbol{S_2} + \beta{S_1^Z}{S_2^Z} \]

${S_1}$ and ${S_2}$ are the spins of the two neighboring atoms in question. We will be considering variants of the special case of $\alpha$ = 0 and $\beta = J$ which is the Ising model of interaction and $\alpha = J$ and $\beta = 0$ which is the Heisenberg model of interaction \cite{Parkinson2010}. 
Spin systems undergo quantum phase transitions. A quantum phase transition is a point of non-analyticity in the energy graph of the ground state of the Hamiltonian of the system caused due to quantum fluctuations at 0 K \cite{atan2018}. Since many complex models can be approximated as spin systems, quantum phase recognition allows us to derive and understand the properties of such systems. 


\begin{figure*}[t]
    \centering
    \begin{subfigure}[b]{\linewidth}
    \begin{minipage}{.1\textwidth}
        \caption{}
        \label{fig:su4-block-circuit}
    \end{minipage}%
    \begin{minipage}{0.90\textwidth}
        \includegraphics[width=\linewidth]{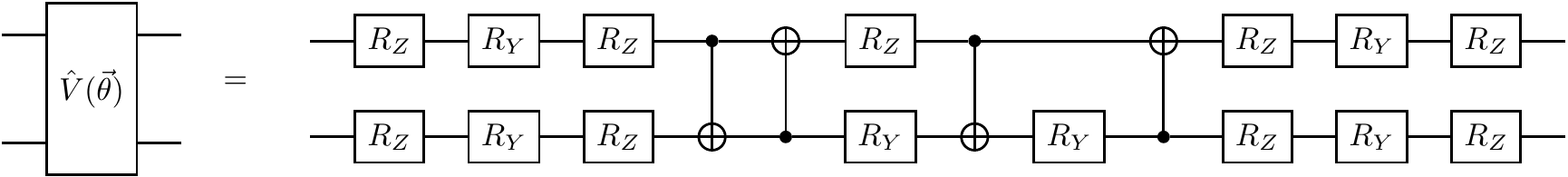}
    \end{minipage}
    \vspace{12pt}
    \end{subfigure}
    
    \begin{subfigure}[b]{\linewidth}
    \begin{minipage}{.1\textwidth}
        \caption{}
        \label{fig:small-block-circuit}
    \end{minipage}%
    \begin{minipage}{0.90\textwidth}
        \includegraphics[width=0.65\linewidth]{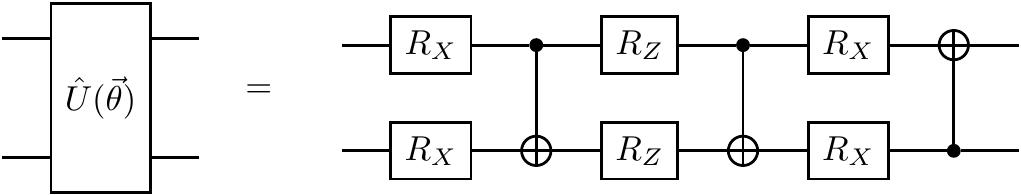}
    \end{minipage}
    \vspace{10pt}
    \end{subfigure}
    \begin{subfigure}[b]{0.48\linewidth}
    \begin{minipage}{.1\textwidth}
        \caption{}
        \label{fig:expressibility-analysis}
    \end{minipage}%
    \begin{minipage}{0.90\textwidth}
        \includegraphics[width=\linewidth]{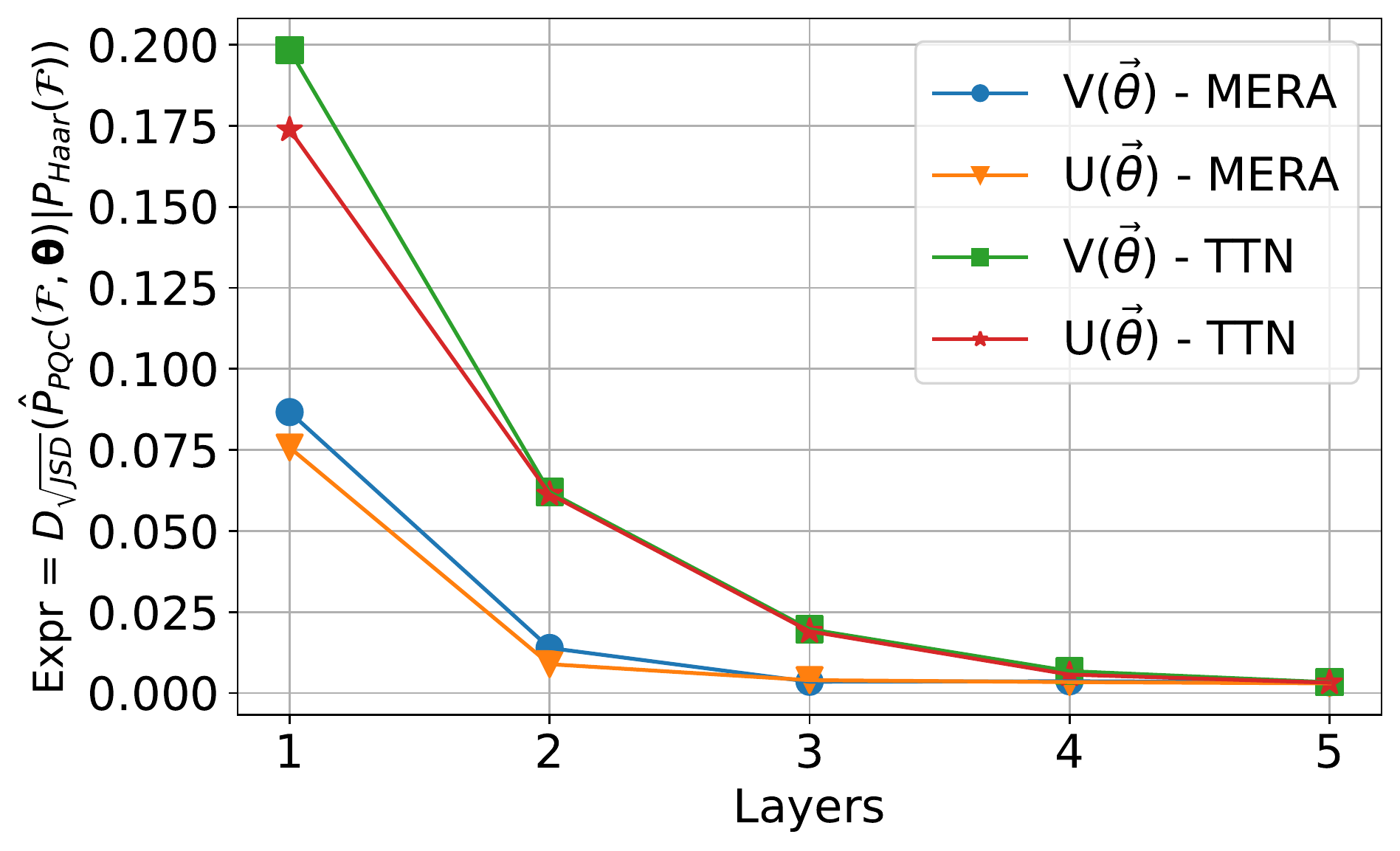}
    \end{minipage}
    \end{subfigure}
    \begin{subfigure}[b]{0.48\linewidth}
    \begin{minipage}{.1\textwidth}
        \caption{}
        \label{fig:entanglingbility-analysis}
    \end{minipage}%
    \begin{minipage}{0.90\textwidth}
        \includegraphics[width=\linewidth]{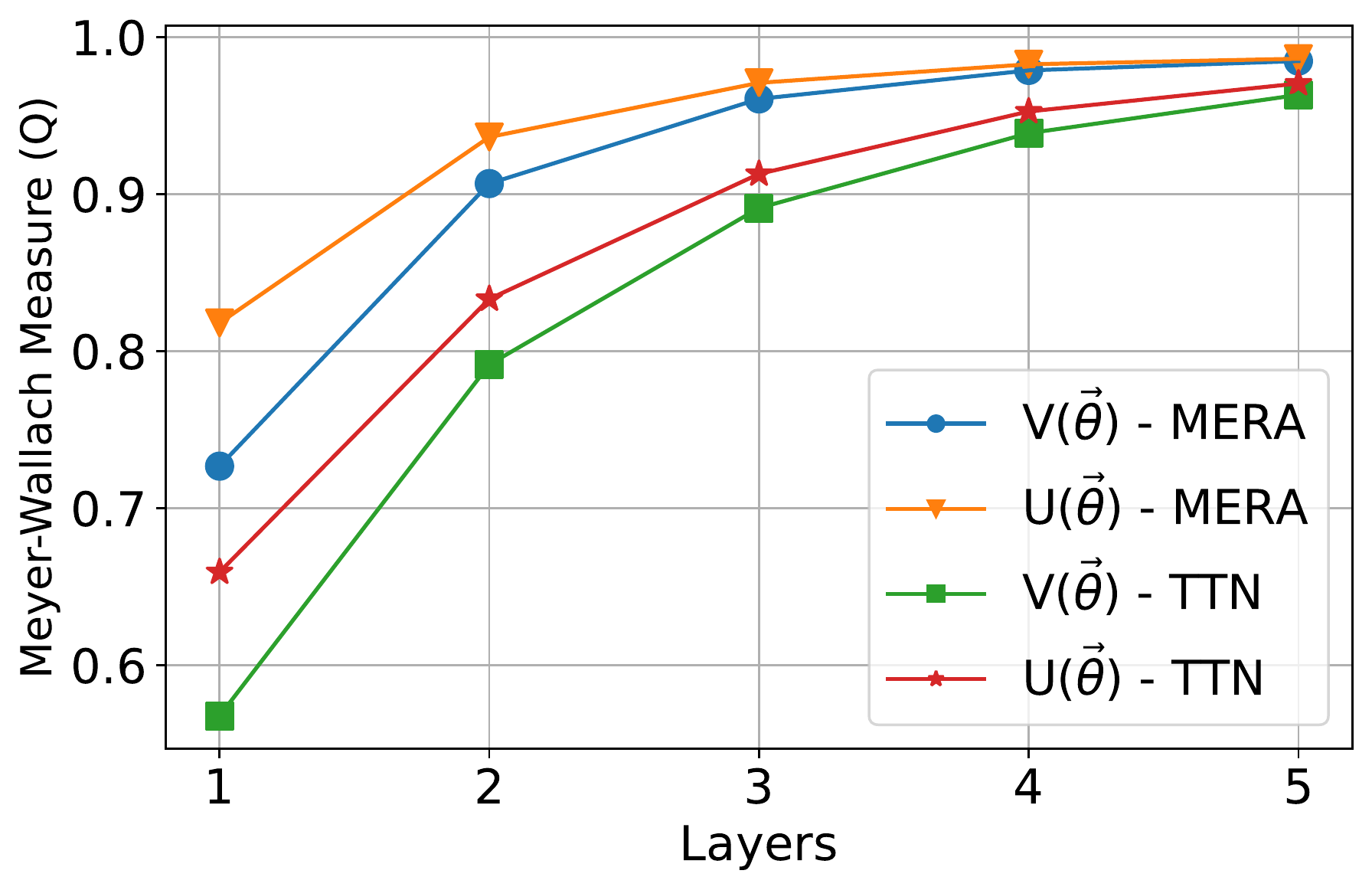}
    \end{minipage}
    \end{subfigure}
    \caption{\textbf{Unitary blocks and their analysis}: The possible choices of unitary blocks for building variational ans\"{a}tze are (a) $V(\vec{\theta})$, which can represent any element from $SU(4)$ group, and (b) $U(\vec{\theta})$, which is a two-qubit entangling unitary. For comparing the effectiveness of built TTN and MERA tensor network ans\"{a}tze, we perform (a) expressibility analysis based on the Jensen-Shannon divergence of fidelity distributions of generated parameterized states with that of Haar states (lower the better), and (b) entangling power analysis based on the Meyer-Wallach measure (higher the better)
    }
    \label{fig:unitary-blocks}
\end{figure*}

\section{\label{sec:setup}Experiments and Results}

\subsection{\label{sec:st:cansatz}Circuit Architecture} 

Our experiments utilize variational quantum circuits based on the tree tensor network (TTN) \cite{Shi2006}, and the multi-scale entanglement renormalization ansatz (MERA) \cite{https://doi.org/10.48550/arxiv.quant-ph/0610099}. The TTN ansatz has a binary-tree-like structure with unitaries applied to the adjacent nodes, as shown in Fig. \ref{fig:ttn-circuit}, which depends on the bond dimension $D$ of the tensor network. As mentioned earlier, the bond dimension equals $D = 2^v$, where $v$ is the number of qubits connecting the subtrees \cite{Bernardi2022}. In our case, we have used $v = 1$; therefore, our ansatz has a bond dimension of two. On the other hand, the structure of the MERA tensor network can be explained using that of TTN itself, where it is constructed by adding a set of unitaries to consecutive nodes of the TTN as shown in Fig. \ref{fig:mera-circuit}. 

\begin{figure*}[!tp]
    \centering
    \includegraphics[width=0.8\linewidth]{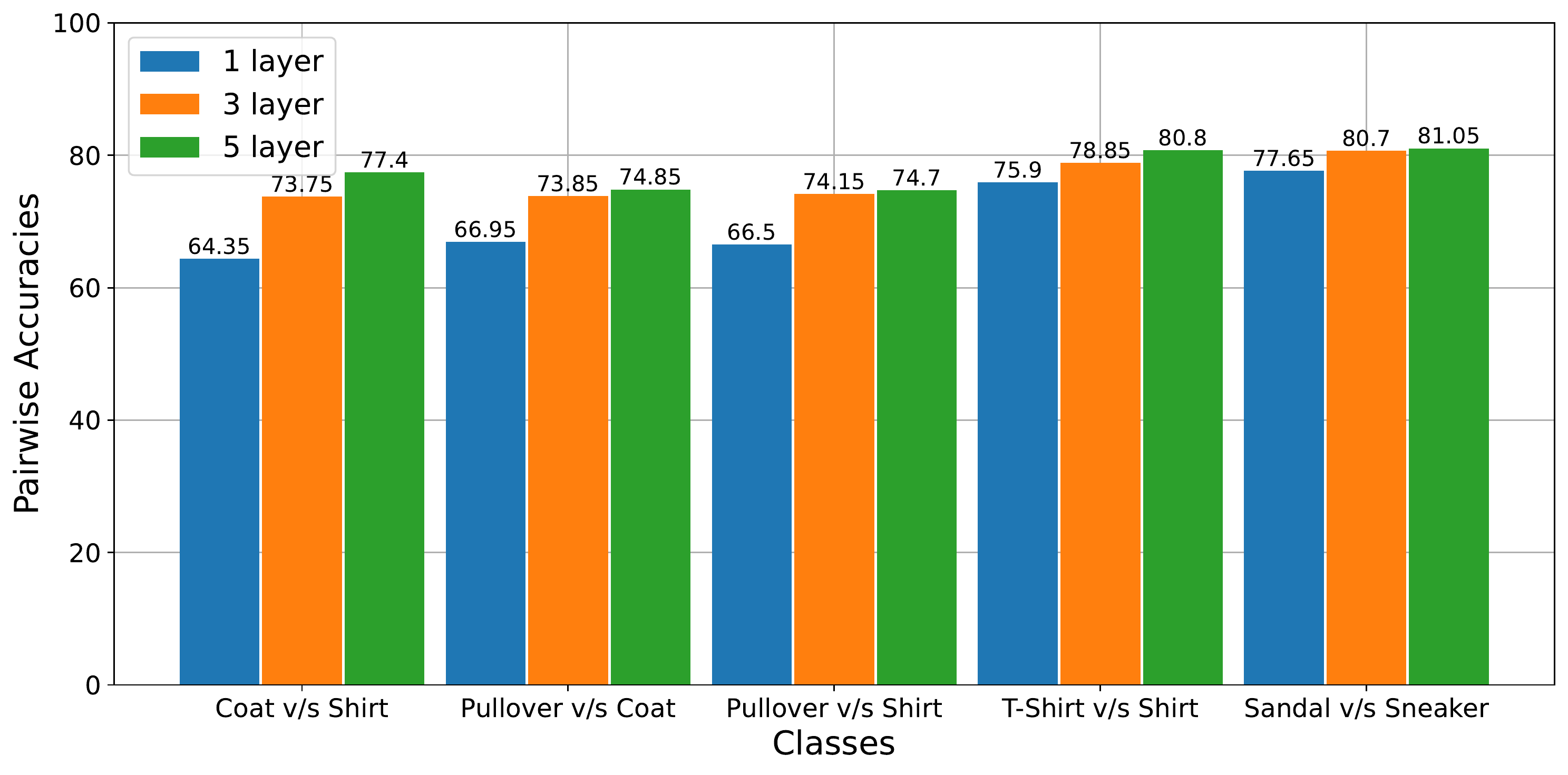}
    \caption{Performance of the MERA tensor network ansatz with different layers on Fashion-MNIST}
    \label{fig:p_acc}
\end{figure*}

\begin{figure}[!tp]
    \centering
    \includegraphics[width=\linewidth]{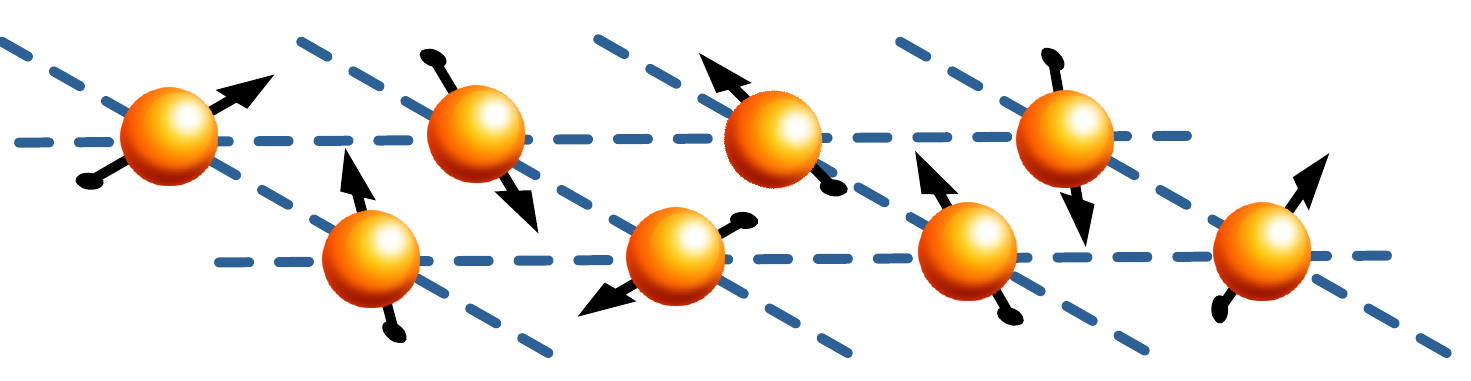}
    \caption{2-D lattice of eight spins in paramagnetic state}
    \label{fig:spins}
\end{figure}

The choice of the ansatz $\mathcal{U}(\vec{\theta})$ is a crucial one, which depends on the unitary block $(U_{B})$ and their bond dimension $(D)$, and results in varied performance between different ansatz structures. In our case, we compare the performance of MERA- and TTN-based ansatz built using the unitary block $\hat{V}(\vec{\theta})$ and $\hat{U}(\vec{\theta})$ based on metrics of expressibility and entangling capability defined in the qLEET library \cite{2022arXiv220502095A}. In particular, we want our ansatz to be more expressive and capable of generating entanglement. For the former, we compare the divergence between the fidelity distributions for the states generated by Haar Random unitaries ($P_{\text{Haar}}(\mathcal{F}, \mathcal{U}_{\text{Haar}})$) and the ansatz ($\hat{P}_{\text{PQC}}(\mathcal{F}, \mathcal{U}(\vec{\theta}))$) using the Jensen-Shannon distance (JSD) \cite{Fuglede2004}, where fidelity $\mathcal{F}(\psi_1, \psi_2) = |\langle \psi_1 | \psi_2 \rangle |^2$ is defined as the squared overlap between the states $\ket{\psi_1}$, $\ket{\psi_2}$ \cite{Jozsa1994} produced by $U_1, U_2 \in \mathcal{U}(\vec{\theta})$ (or $\mathcal{U}_{\text{Haar}}$). We use this to define \textit{expressivity} (Expr $\in [0, 1]$) of the ans\"{a}tze as given below:\vspace{-0.4pt}
\begin{equation}
    \text{Expr} = D_{\sqrt{JSD}}(\hat{P}_{PQC}(\mathcal{F}, \mathcal{U}(\vec{\theta})) | P_{\text{Haar}}(\mathcal{F}, \mathcal{U}_{\text{Haar}}),
\end{equation}
The smaller this distance, i.e., divergence, the closer ansatz is to Haar random unitaries and hence more expressive it comes out to be. In contrast, to compare the latter, we use an entanglement measure known as the Mayer-Wallach measure ($Q$) \cite{Meyer2002}, which quantifies the average entanglement in all the states produced by an ansatz by measuring the average linear entropy over all possible single-qubit subsystems (Eq. \ref{eq:mayer-wallach}).
\begin{equation}
    \label{eq:mayer-wallach}
	Q = \frac{2}{|\vec{\theta}|}\sum_{\theta_{i}\in \vec{\theta}}\Bigg(1-\frac{1}{n}\sum_{k=1}^{n}\text{Tr}(\rho_{k}^{2}(\theta_{i}))\Bigg),\quad 0\leq Q\leq 1
\end{equation}
where $n$ is the total number of qubits, $\rho (\vec{\theta}) = \ket{\psi(\vec{\theta}}\bra{\psi(\vec{\theta})}$ is the density matrix for the parameterized pure state $\ket{\psi(\theta)}$ and $\rho_k (\vec{\theta})$ is the reduced single-qubit density matrix for the $k^{\text{th}}$ qubit after tracing out the rest. For any given ansatz $\mathcal{U}(\vec{\theta})$, the larger the value of $Q$ is, the more capable in general it would be to produce entanglement between qubits, i.e., more entangled states.

We present the structure description of $U(\vec{\theta})$ and $V(\vec{\theta})$ in Figs. \ref{fig:su4-block-circuit} and \ref{fig:small-block-circuit}. The first one is a general element of the $SU(4)$ group, which can be decomposed into four controlled-NOTs and 15 single-qubits rotations \cite{Lazzarin2022}. In distinction, the other one is a two-qubit entangler gate comprising three controlled-NOTS and six single-qubit rotations arranged in a layer-wise manner. 

In the Figs. \ref{fig:expressibility-analysis} and \ref{fig:entanglingbility-analysis}, we look at expressibility and entangling capability, respectively. As a general trend, we see that MERA-based ansatz is more expressible and generates more entangled states than TTN-based ansatz. Additionally, amongst $U(\vec{\theta})$ and $V(\vec{\theta})$, in both the cases, for single-layered circuits ($L=1$), the ansatz built using $U(\vec{\theta})$ comes out to be more effective than $V(\vec{\theta})$. Moreover, since the number of variational parameters is lesser in case of the $U(\vec{\theta})$ block as compared to the $V(\vec{\theta})$ block, with a ratio of $2:5$, it will be easier to optimize the former and hence it will be more scalable for larger systems. Additionally, the circuits with $U(\vec{\theta})$ and $V(\vec{\theta})$ blocks become equally expressible for both MERA- and TTN- based ansatz for multiple layers ($L>1$). However, the $U(\vec{\theta})$ block produces more entangled states than the $V(\vec{\theta})$ block. Therefore, based on these observations, we have used tensor network ansatz based on the $U_{B} = U(\vec{\theta})$. 

\definecolor{high}{HTML}{2E8B57}  
\definecolor{low}{HTML}{FFFF33}  
\newcommand*{\opacity}{100}
\newcommand*{\minval}{0.6}
\newcommand*{\maxval}{1.0}
\newcommand{\gradient}[1]{
    \ifdimcomp{#1pt}{>}{\maxval pt}{#1}{
        \ifdimcomp{#1pt}{<}{\minval pt}{#1}{
            \pgfmathparse{int(round(100*(#1/(\maxval-\minval))-(\minval*(100/(\maxval-\minval)))))}
            \xdef\tempa{\pgfmathresult}
            \cellcolor{high!\tempa!low!\opacity} #1
    }}
}
\newcommand{\gradientcell}[6]{
    \ifdimcomp{#1pt}{>}{#3 pt}{#1}{
        \ifdimcomp{#1pt}{<}{#2 pt}{#1}{
            \pgfmathparse{int(round(100*(#1/(#3-#2))-(\minval*(100/(#3-#2)))))}
            \xdef\tempa{\pgfmathresult}
            \cellcolor{#5!\tempa!#4!#6} #1
    }}
}
\begin{table*} 
        \centering
        \Huge
        \resizebox{0.75\linewidth}{!}{%
            \begin{tabular}{|*{12}{c|}}
              \cline{1-2} 
              \multicolumn{1}{|c|}{T-shirt} & \multicolumn{1}{c|}{-}  \\
              
              \cline{3-3} 
              \multicolumn{1}{|c|}{Trouser} & \multicolumn{1}{c|}{\gradient{0.953}}  & \multicolumn{1}{c|}{-} \\
              
              \cline{4-4} 
              {Pullover} & \gradient{0.894} & \gradient{0.966} & \multicolumn{1}{c|}{-} \\
              
              \cline{5-5} 
              {Dress} & \gradient{0.8655} & \gradient{0.9155} & \gradient{0.9615} & \multicolumn{1}{c|}{-} \\
              
              \cline{6-6} 
              {Coat} & \gradient{0.8315} & \gradient{0.944} & \gradient{0.6695} & \gradient{0.8875} & \multicolumn{1}{c|}{-} \\
              
              \cline{7-7} 
              {Sandal} & \gradient{0.9085} & \gradient{0.977} & \gradient{0.9745} & \gradient{0.9795} & \gradient{0.8925} & \multicolumn{1}{c|}{-} \\
              
              \cline{8-8}
              {Shirt} & \gradient{0.759} & \gradient{0.9405} & \gradient{0.665} & \gradient{0.8935} & \gradient{0.6435} & \gradient{0.9725} & \multicolumn{1}{c|}{-} \\
              
               \cline{9-9} 
              {Sneaker} & \gradient{0.988} & \gradient{0.992} & \gradient{0.9925} & \gradient{0.993} & \gradient{0.995} & \gradient{0.7765} & \gradient{0.994} & \multicolumn{1}{c|}{-} \\
    
              \cline{10-10} 
              {Bag} & \gradient{0.9155} & \gradient{0.9645} & \gradient{0.939} & \gradient{0.9445} & \gradient{0.8975} & \gradient{0.7895} & \gradient{0.9335} & \gradient{0.9215} & \multicolumn{1}{c|}{-} \\  
              
              \cline{11-11} 
              {Ankle boot} & \gradient{0.9845} & \gradient{0.9815} & \gradient{0.993} & \gradient{0.98} & \gradient{0.98} & \gradient{0.79} & \gradient{0.9845} & \gradient{0.8925} & \gradient{0.9895} & \multicolumn{1}{c|}{-} \\ 
              
              \cline{1-11} 
              \multicolumn{1}{|c|}{Layer: 1}  & {T-shirt} & Trouser & Pullover & Dress & Coat & Sandal & Shirt & Sneaker & Bag & \multicolumn{1}{c|}{Ankle boot} \\
              
              \cline{1-11}
             
            \end{tabular}
        }
         \caption{Pairwise accuracy on the classes of the Fashion MNIST dataset for one layer of the MERA tensor network}
        \label{Tab:1_acc}
        
        \Huge
        \begin{center}
        \resizebox{0.75\linewidth}{!}{%
            \begin{tabular}{|*{12}{c|}}
              \cline{1-2}
              \multicolumn{1}{|c|}{T-shirt} & \multicolumn{1}{c|}{-} \\
              
              \cline{3-3}
              {Trouser} & \gradient{0.9585}  & \multicolumn{1}{c|}{-} \\
              
              \cline{4-4} 
              {Pullover} & \gradient{0.9375} & \gradient{0.974} & \multicolumn{1}{c|}{-} \\
              
              \cline{5-5} 
              {Dress} & \gradient{0.8805} & \gradient{0.9445} & \gradient{0.9655} & \multicolumn{1}{c|}{-} \\
              
              \cline{6-6} 
              {Coat} & \gradient{0.8955} & \gradient{0.959} & \gradient{0.7385} & \gradient{0.891} & \multicolumn{1}{c|}{-}  \\
              
              \cline{7-7} 
              {Sandal} & \gradient{0.982} & \gradient{0.986} & \gradient{0.989} & \gradient{0.9915} & \gradient{0.98} & \multicolumn{1}{c|}{-} \\
              
              \cline{8-8} 
              {Shirt} & \gradient{0.7875} & \gradient{0.963} & \gradient{0.7415} & \gradient{0.9005} & \gradient{0.7375} & \gradient{0.9825} & \multicolumn{1}{c|}{-} \\
              
               \cline{9-9} 
              {Sneaker} & \gradient{0.992} & \gradient{0.9945} & \gradient{0.9955} & \gradient{0.9965} & \gradient{0.998} & \gradient{0.807} & \gradient{0.9955} & \multicolumn{1}{c|}{-} \\
    
              \cline{10-10} 
              {Bag} & \gradient{0.968} & \gradient{0.9775} & \gradient{0.958} & \gradient{0.962} & \gradient{0.973} & \gradient{0.94} & \gradient{0.9475} & \gradient{0.964} & \multicolumn{1}{c|}{-} \\  
              
              \cline{11-11} 
              {Ankle boot} & \gradient{0.989} & \gradient{0.986} & \gradient{0.998} & \gradient{0.9915} & \gradient{0.992} & \gradient{0.799} & \gradient{0.992} & \gradient{0.8975} & \gradient{0.9915} & \multicolumn{1}{c|}{-} \\ 
              
              \cline{1-11} 
              \multicolumn{1}{|c|}{Layers: 3}  & {T-shirt} & Trouser & Pullover & Dress & Coat & Sandal & Shirt & Sneaker & Bag & \multicolumn{1}{c|}{Ankle boot} \\
              
              \cline{1-11}
             
            \end{tabular}
        }
        \caption{Pairwise accuracy on the classes of the Fashion MNIST dataset for three layers of the MERA tensor network}
        \label{Tab:3_acc}
        \end{center}

        \Huge
        \begin{center}
        \resizebox{0.75\linewidth}{!}{%
            \begin{tabular}{|*{12}{c|}}
              \cline{1-2} 
              \multicolumn{1}{|c|}{T-shirt} & \multicolumn{1}{c|}{-}  \\
              
              \cline{3-3} 
              {Trouser} & \gradient{0.959}  & \multicolumn{1}{c|}{-} \\
              
              \cline{4-4} 
              {Pullover} & \gradient{0.939} & \gradient{0.9755} & \multicolumn{1}{c|}{-} \\
              
              \cline{5-5} 
              {Dress} & \gradient{0.8895} & \gradient{0.9515} & \gradient{0.9655} & \multicolumn{1}{c|}{-} \\
              
              \cline{6-6}
              {Coat} & \gradient{0.9285} & \gradient{0.964} & \gradient{0.7485} & \gradient{0.8965} & \multicolumn{1}{c|}{-}  \\
              
              \cline{7-7} 
              {Sandal} & \gradient{0.99} & \gradient{0.9915} & \gradient{0.995} & \gradient{0.992} & \gradient{0.9935} & \multicolumn{1}{c|}{-} \\
              
              \cline{8-8} 
              {Shirt} & \gradient{0.788} & \gradient{0.964} & \gradient{0.747} & \gradient{0.905} & \gradient{0.774} & \gradient{0.9865} & \multicolumn{1}{c|}{-} \\
              
               \cline{9-9} 
              {Sneaker} & \gradient{0.9925} & \gradient{0.9975} & \gradient{0.999} & \gradient{0.9975} & \gradient{0.9985} & \gradient{0.8105} & \gradient{0.997} & \multicolumn{1}{c|}{-}  \\
    
              \cline{10-10}
              {Bag} & \gradient{0.9695} & \gradient{0.9775} & \gradient{0.963} & \gradient{0.969} & \gradient{0.974} & \gradient{0.9425} & \gradient{0.955} & \gradient{0.9805} & \multicolumn{1}{c|}{-} \\  
              
              \cline{11-11} 
              {Ankle boot} & \gradient{0.99} & \gradient{0.9885} & \gradient{0.9985} & \gradient{0.9945} & \gradient{0.992} & \gradient{0.8315} & \gradient{0.9965} & \gradient{0.898} & \gradient{0.994} & \multicolumn{1}{c|}{-}  \\ 
              
              \cline{1-11} 
              \multicolumn{1}{|c|}{Layers: 5}  & {T-shirt} & Trouser & Pullover & Dress & Coat & Sandal & Shirt & Sneaker & Bag & \multicolumn{1}{c|}{Ankle boot}  \\
              
              \cline{1-11}
             
            \end{tabular}
        }
        \caption{Pairwise accuracy on the classes of the Fashion MNIST dataset for five layers of the MERA tensor network}
        \label{Tab:5_acc}
        \end{center}
        
    \end{table*} 
\normalsize

\subsection{\label{sec:st:tasks}Tasks}

\subsubsection{\label{sec:st:imgclass}Image Classification}

\subsubsubsection{\label{sec:st:dataset}Dataset}
We have conducted the image classification tasks on the Fashion-MNIST dataset \cite{https://doi.org/10.48550/arxiv.1708.07747}. It is a set of $28\times28$ grayscale images with $60,000$ train and $10,000$ test samples spread uniformly among 10 classes (tshirt, trousers, pullover, etc.) In our experiments, the train set was split in a ratio of $5:1$ into balanced train and validation sets. Therefore, our data was split into the train, validation, and test classes in the ratio $5:1:1$, with each class distributed uniformly in each of these sets.

Most of the image classification tasks done using quantum tensor networks use the MNIST dataset \cite{lecun-mnisthandwrittendigit-2010}. We have chosen the Fashion-MNIST dataset because it is less explored in the quantum machine learning literature and is more complicated than the MNIST dataset, which is essentially solved at this point \cite{benchmark_dashboard}. Therefore obtaining better accuracies at Fashion-MNIST would represent the effectiveness of the learning models better.

\subsubsubsection{\label{sec:st:enc}Encoding Strategy}
In order to process classical data using a quantum circuit, we first need to embed it in a quantum state. For our experiments, we have used amplitude embedding in which encoding an image of size $N\times M$ will require $log_2(N \times M)$ qubits. Each image of size $28\times 28$ is first converted to a linear vector of size $1 \times {28^2}$. In case the image size is too large to process, it is first resized and then transformed into the image vector. The image vector is then mapped to a state in the Hilbert Space. A variety of feature maps can be used for this purpose. So each image in the Fashion-MNIST dataset was first resized and converted to an image vector. The image vector was then normalized and encoded into the amplitudes of an eight qubit quantum state.


\begin{table*}[t]
\centering
\resizebox{0.9\linewidth}{!}{%
\rowcolors{4}{white}{yellow!25}
\begin{tabular}{|c|c|c|c|c|c|}
\hline
\multirow{2}{*}{\textbf{Spin}} &
  \multirow{2}{*}{\textbf{Lattice}} &
  \multirow{2}{*}{\textbf{Tensor Network}} &
  \multicolumn{3}{c|}{\textbf{Test Accuracies}} \\
   \textbf{Models}
   & 
   & \textbf{States}
   &
  \begin{tabular}[c]{@{}c@{}}\textbf{8 spins} \\ \textbf{(simulator)}\end{tabular} &
  \begin{tabular}[c]{@{}c@{}}\textbf{4 spins} \\ \textbf{(simulator)}\end{tabular} &
  \begin{tabular}[c]{@{}c@{}}\textbf{4 spins} \\ \textbf{(IBMQ Nairobi)}\end{tabular}\\ \hline
XXZ-HM       & 1-D   & MERA & 98.6 $\pm$ 0.70  & 98.6 $\pm$ 0.38 & 74.0   \\
XXZ-HM       & 1-D   & TTN  & 96.5 $\pm$ 1.03 & 98.5 $\pm$ 0.32 & 72.6 \\
TFIM         & 1-D   & MERA & 99.8 $\pm$ 0.06 & 98.6 $\pm$ 0.08 & 84.5 \\
TFIM         & 1-D   & TTN  & 98.3 $\pm$ 0.10  & 99.0 $\pm$ 0.03 & 86.2 \\
XXZ-HM       & 2-D & MERA & 98.5 $\pm$ 0.88 & 98.4 $\pm$ 0.27 & 68.6 \\
XXZ-HM       & 2-D & TTN  & 96.3 $\pm$ 1.25 & 98.1 $\pm$ 0.31 & 64.0 \\
TFIM         & 2-D & MERA & 99.8 $\pm$ 0.06 & 98.8 $\pm$ 0.18 & 72.1 \\
TFIM         & 2-D & TTN  & 98.0 $\pm$ 0.09   & 99.1 $\pm$ 0.13 & 73.6\\\hline
\end{tabular}%
}
\caption{Performance of the TTN and MERA tensor networks on recognizing correct phases of various XXZ Heisenberg (XXZ-HM) and transverse-field Ising (TFIM) spin systems on one-dimensional (linear) and two-dimensional (rectangular) lattices. For eight spin systems, simulations were performed numerically on a quantum simulator, and results were averaged over five trials. Whereas for the four spins systems, along with similar numerical simulations, experiments were also executed on the IBMQ Nairobi (\textit{imbq\_nairobi}) \cite{IBMQ}, a seven-qubit quantum hardware, and the best results out of three trials are being reported here.}
\label{Tab:tn_acc}
\end{table*}

\subsubsubsection{\label{sec:st:opt}Optimization and Hyperparameters}
The ADAM optimizer \cite{https://doi.org/10.48550/arxiv.1412.6980} was used to optimize the training process with a learning rate of $0.01$. A mini-batch size of $20$ was used, and the model was trained over $40$ epochs to minimize the cross-entropy loss \cite{10.5555/3327546.3327555}. Computation of both the loss and pair-wise accuracies was done by using $\langle Z_3 \rangle$ obtained by computational basis measurement of the third qubit.

\subsubsubsection{Results}
We used eight qubit ansatz based on both TTN and MERA tensor network states. Amongst them, the latter obtained reasonably better results for all pairs of classes of the Fashion MNIST dataset and therefore, we present only its results here.  Predictably, we see better performance on classes that are more unlike each other, like Pullover vs. Ankle boot ($99.3\%$), than classes that are similar to each other, like coat and shirt ($64.35\%$).

Increasing the number of layers to three and five has shown an increase in the pairwise accuracy, especially in our previous case of the coat and shirt labels where the accuracy increases from $64.35\%$ to $73.75\%$ to $77.4\%$ as can be seen in Tables \ref{Tab:1_acc}, \ref{Tab:3_acc} and \ref{Tab:5_acc}.
Such a trend is observed in most classes where single-layered tensor network ans\"{a}tze did not perform very well. 

We observe a bigger difference in accuracy when going from one layer to three layers than when going from three layers to five layers. In fact, in most cases, we see a similar performance in ans\"{a}tze with three and five layers. Fig. \ref{fig:p_acc} shows the pairwise accuracy for certain pairs of classes as the number of layers increases, which is corroborated by the increasing trends of expressibility and entangling power in Figs. \ref{fig:expressibility-analysis} and \ref{fig:entanglingbility-analysis}.

\subsubsection{\label{sec:st:phaseRecog}Quantum Phase Recognition} 

\subsubsubsection{\label{sec:st:datagen}Models and Data generation} 

\subsubsubsubsection{\label{sec:st:1-DIsing}1-D Transverse-field Ising Model} 
The transverse-field Ising model in one dimension is characterized by the following Hamiltonian:
\begin{equation}
  \hat{H}(h) = J\sum_{i=1}^{n} {\hat{\sigma}_i^z} {\hat{\sigma}_{i+1}^z} + h\sum_{i=1}^{n} {\hat{\sigma}_i^x},  
\end{equation}
where $J$ is the coupling constant, $h$ the external magnetic field and $\sigma_{i}^z$ and $\sigma_{i}^x$ represents the Pauli matrices $Z$ and $X$ acting on the $i^{th}$ spin. We have taken $J = 1$ in our experiments, so when $h < 1$, the nearest-neighbor term dominates. This leads to the spins aligning in either an up or down direction, resulting in a disordered paramagnetic phase. For $h > 1$, the second term dominates, and the spins end up aligning themselves with the external magnetic field leading to an ordered ferromagnetic phase. A phase transition for this system between these two phases occurs at $h = J$ \cite{atan2018}.

In our experiments, we have generated $1000$ ground states for linear chain systems with four and eight spins using the given Hamiltonian with $J=1$ and $h$ varying from $0$ to $2J$.

\begin{figure*}[!t]
    \centering
    \begin{subfigure}[b]{0.48\linewidth}
    \begin{minipage}{.1\textwidth}
        \caption{}
        \label{fig:ising-pred-prob-sim}
    \end{minipage}%
    \begin{minipage}{0.90\textwidth}
        \includegraphics[width=\linewidth]{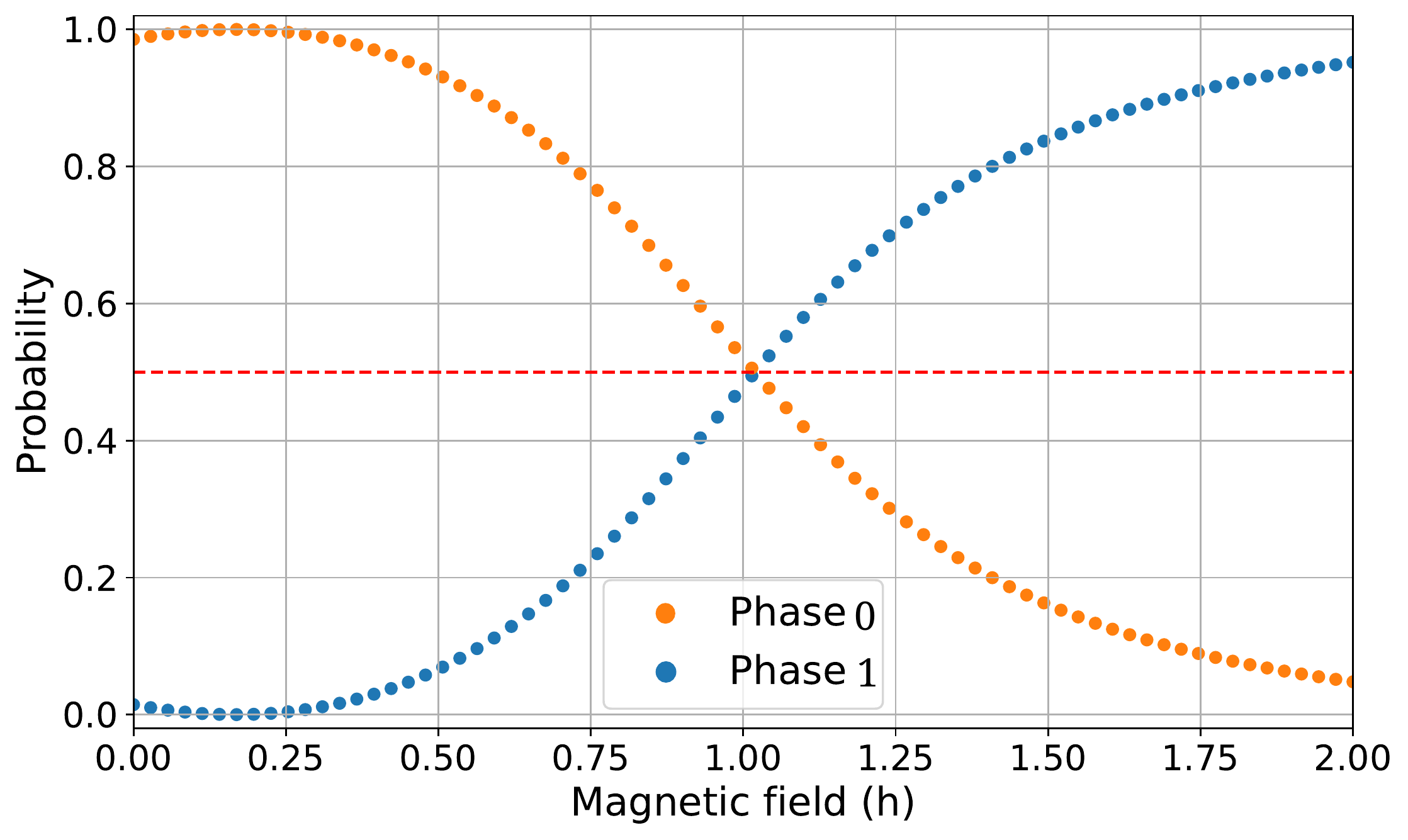}
    \end{minipage}
    \end{subfigure}
    \begin{subfigure}[b]{0.48\linewidth}
    \begin{minipage}{.1\textwidth}
        \caption{}
        \label{fig:ising-pred-prob-hardware}
    \end{minipage}%
    \begin{minipage}{0.90\textwidth}
        \includegraphics[width=\linewidth]{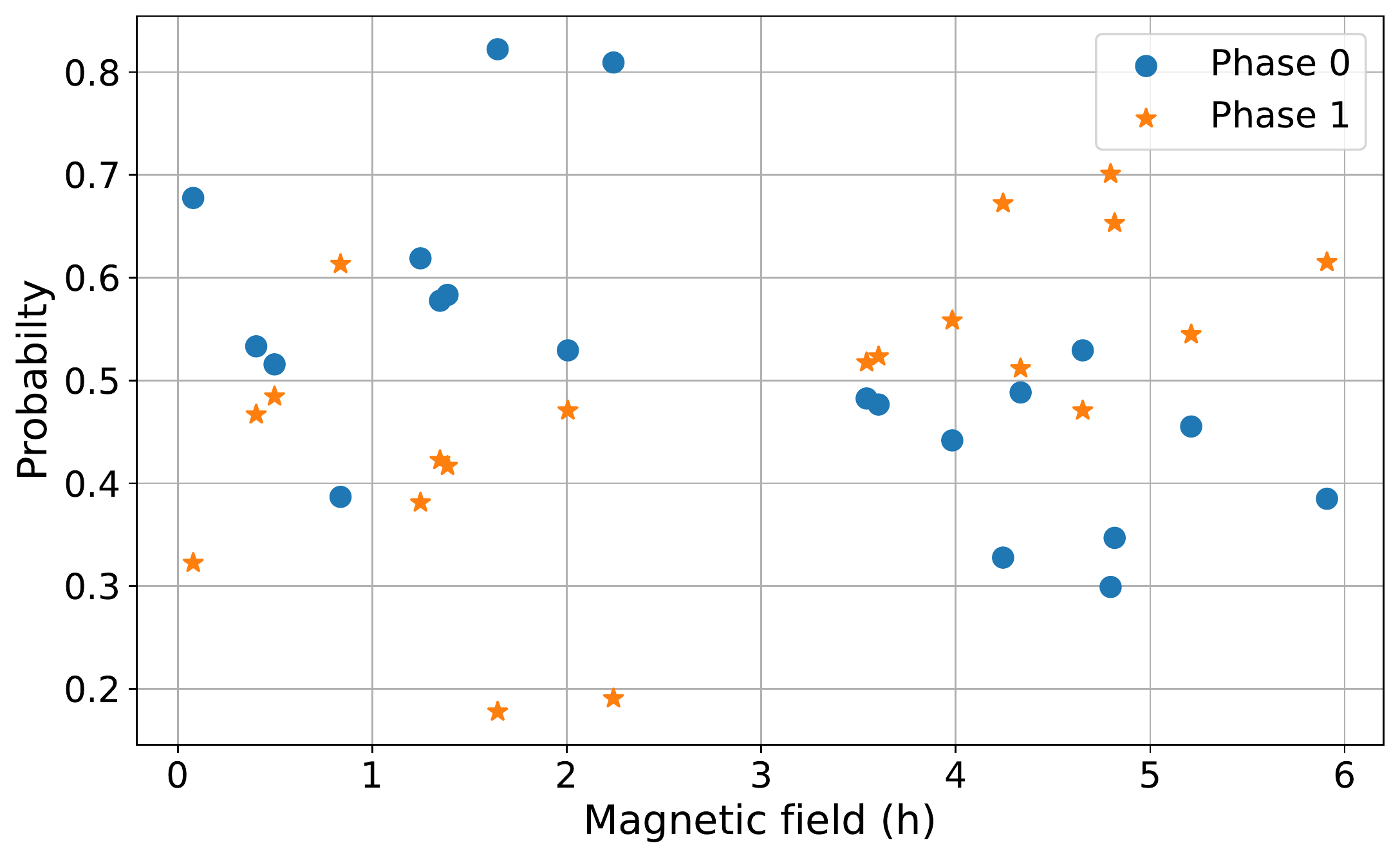}
    \end{minipage}
    \end{subfigure}
    \caption{\textbf{Prediction probabilities of phases with MERA based ansatz}: (a) for transverse-field Ising model in 1-D case (noiseless simulation), and (b) For transverse-field Ising model in 2-D case (executed on IBMQ Nairobi (\textit{ibmq\_nairobi}), 7-qubit hardware \cite{IBMQ})
    }
    \label{fig:pred-probs-phases}
\end{figure*}

\subsubsubsubsection{\label{sec:st:2-DIsing}2-D Transverse-field Ising Model} 
The two-dimensional transverse-field Ising model has the same Hamiltonian as the one-dimensional case. However, a phase transition is observed at $h \approx 3.01J$ \cite{Hashizume2022}.  The data is generated for 1000 points for 2-D lattices with four spins $(2 \times 2)$ and eight spins $(2 \times 4)$ using the given Hamiltonian with $J=1$ and $h$ varying from $0$ to $6J$. The phase transition is seen at $h = 3.01J$. Fig. \ref{fig:spins} shows a 2-D lattice $(2 \times 4)$ of spin systems.

\subsubsubsubsection{\label{sec:st:1-DHeisenberg}1-D XXZ Heisenberg Model} The one-dimensional XXZ Heisenberg model is described using the following Hamiltonian:

\begin{equation}
    \hat{H}(h) = J\left[\sum_{i=1}^{n} {\hat{\sigma}_{i}^x}{\hat{\sigma}_{i+1}^x} + {\hat{\sigma}_i^y}{\hat{\sigma}_{i+1}^y} + \Delta{\hat{\sigma}_i^z}{\hat{\sigma}_{i+1}^z}\right]
\end{equation}

Where $J$ is again a coupling constant, taken as 1 in our experiments, and $\Delta$ introduces anisotropy in the interaction along the $\hat{z}$-direction. It is observed that for $\Delta \rightarrow  \infty$, the system is in the antiferromagnetic/N\'eel state, i.e., all spins are alternating spin-up or spin-down. As $\Delta \rightarrow 1$, the spins begin to reorient themselves, and for $-1  < \Delta < 1$, they remain in the $\hat{x}-\hat{y}$ plane, putting the system in a paramagnetic phase \cite{Parkinson2010}. Finally, for $\Delta < -1$, all the spins arrange themselves in the same direction resulting in a ferromagnetic phase. 
Therefore, for $J = 1$, phase transitions happen clearly at $\Delta = 1$ and $\Delta = -1$ \cite{Parkinson2010, Franchini2017}. 

The 1-D XXZ Heisenberg Model data was generated for linear chain systems with four spins $(1 \times 4)$ and eight spins $(1 \times 8)$ using the given Hamiltonian with $J=1$ and $\Delta$ varying from $-2$ to $2$ for $1000$ points.

\subsubsubsubsection{\label{sec:st:2-DHeisenberg}2-D XXZ Heisenberg Model}
The two-dimensional XXZ Heisenberg model has the same Hamiltonian as the one-dimensional case with a phase transition occurring at $\Delta =$ 1 and -1 as well \cite{Saryer2019}. The data generation process remains the same as in the one-dimensional case.

\subsubsubsection{\label{sec:st:training}Training}
A quantum circuit was trained using the variational quantum algorithm as shown in Fig \ref{fig:mera-circuit}. The data was split randomly into the train, validation, and test sets in the ratio $3:1:1$, with each set having a balanced distribution of the classes. For all the experiments, measurements were taken on two readout qubits $(q_i, q_j)$. These were the second and third qubits for the four-spin systems and the third and sixth qubits for the eight-spin systems. For both Ising and Heisenberg models, we calculate expectation values $\langle Z_i\rangle$, $\langle Z_j\rangle$ and $\langle Z_i Z_j\rangle$ to compute probabilities of the elements of computational basis corresponding to each class (phase) inspired by the amplitude decoding method introduced in \cite{Lazzarin2022}. These probabilities were fed to a softmax function for normalization, whose outputs were used to calculate the cross-entropy loss \cite{10.5555/3327546.3327555}. A batch size of $8$ was used with a learning rate of $0.002$ for the MERA-based ansatz and $0.0008$ for the TTN-based ansatz. The ADAM optimizer \cite{https://doi.org/10.48550/arxiv.1412.6980} was used to optimize the training process over 2000 iterations.

\subsubsubsection{Results}

We see that both MERA- and TTN-based ans\"{a}tze perform well for both four-spin and eight-spin systems, with overall better performance for the transverse-field Ising models than the Heisenberg models (Table \ref{Tab:tn_acc}). Moreover, their performance in the 1-D cases is better than in the more complicated 2-D case. In Fig. \ref{fig:ising-pred-prob-sim}, we show the results outputted by our model for the transverse-field Ising model with eight spins on a linear (chain) lattice. When the probability of a phase is more than $50\%$, we assign our output the label corresponding to that phase. We see that our model is more confident when the value of h is further from the point of phase transition, i.e., when $h$ is 1. Similarly, Fig. \ref{fig:ising-pred-prob-hardware} shows the values of our model's outputs when the models were trained and executed on the IBMQ Nairobi (\textit{ibmq\_nairobi}) \cite{IBMQ}, which is a 7-qubit superconducting quantum hardware for the transverse-field Ising model with four spins on a square $(2\times 2)$ lattice.


\begin{figure*}[t]
    \centering
    \begin{subfigure}[b]{0.30\linewidth}
    \hspace{-25pt}
    \begin{minipage}{.08\textwidth}
        \caption{}
        \label{fig:4-spin-ttn}
    \end{minipage}%
    \begin{minipage}{0.90\textwidth}
    \includegraphics[width=.92\linewidth]{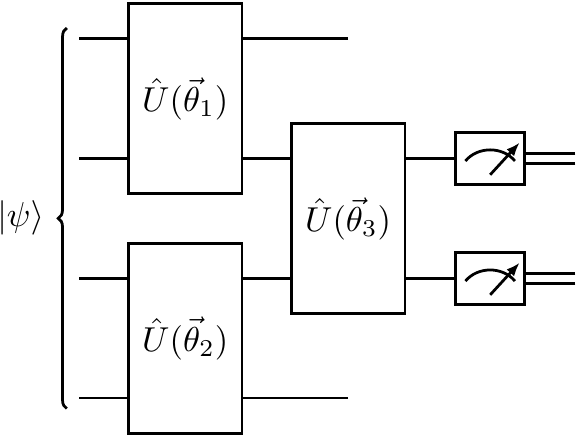}
    \end{minipage}
    \end{subfigure}
    \begin{subfigure}[b]{0.32\linewidth}
    \hspace{-25pt}
    \begin{minipage}{.1\textwidth}
        \caption{}
        \label{fig:4-spin-mera}
    \end{minipage}%
    \begin{minipage}{0.98\textwidth}
        \includegraphics[width=\linewidth]{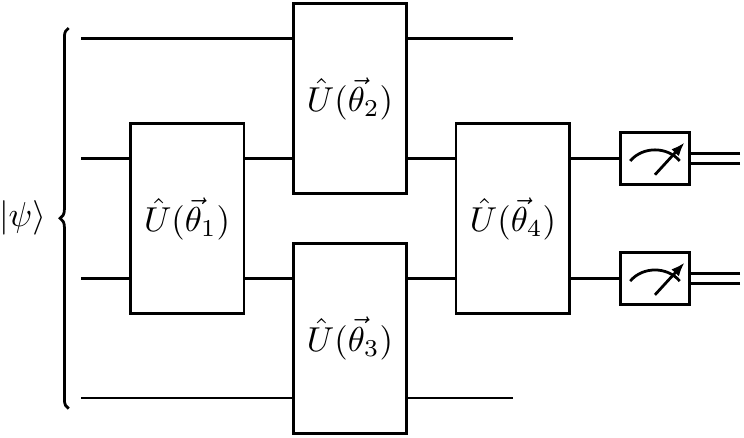}
    \end{minipage}%
    \end{subfigure}
    \begin{subfigure}[b]{0.32\linewidth}
    \hspace{-15pt}
    \begin{minipage}{.1\textwidth}
        \caption{}
        \label{fig:4-spin-mera-mod}
    \end{minipage}%
    \begin{minipage}{0.98\textwidth}
        \includegraphics[width=\linewidth]{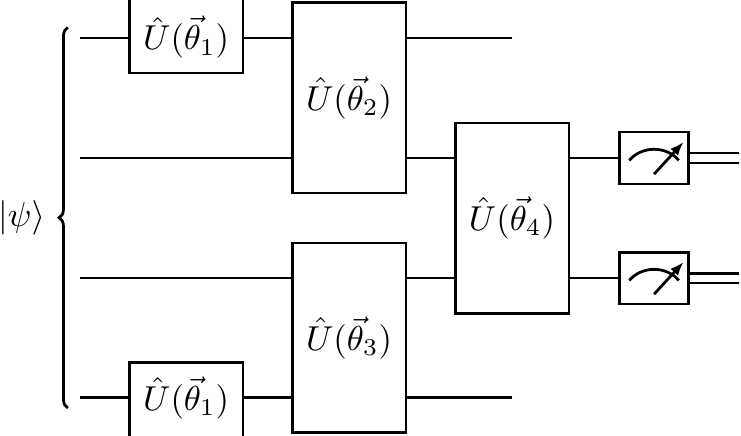}
    \end{minipage}
    \end{subfigure}
    \caption{\textbf{Tensor network ans\"{a}tze for four-spin systems}: Structures of variational ans\"{a}tze based on the (a) tree tensor network (TTN) and the (b) multi-scale entanglement renormalization ansatz (MERA) tensor network. (c) Modified structure of MERA tensor network ansatz with changed first unitary block}
    \label{fig:4-spin-ttn-mera}

\end{figure*}

\section{\label{sec:conclusions}Discussions and Conclusions}

In this paper, we have studied the performance of quantum tensor networks for image classification tasks and quantum phase recognition tasks of spin systems. 

We have extended the previous works done in this domain in the following two ways. First, we have presented a strategy based on metrics like expressibility and entangling capability of the parameterized circuits to choose a well-suited block structure for the tensor-network inspired ans\"{a}tze. Such analysis was corroborated by the results obtained for the image classification task, where we were able to increase the performance of our classifiers by increasing the number of layers of the circuits. Second, for the quantum phase recognition task, we have attempted to study spin systems on 2-D lattices with the tensor-network ansatz, which are generally more challenging than those on 1-D spin lattices that have been studied in the literature until now \cite{Lazzarin2022, PhysRevA.102.012415}.

In the image classification task, the pairwise accuracies between the different classes of the Fashion-MNIST dataset were calculated for 1, 3, and 5 layers of the MERA tensor network ansatz. The results are shown in Tables \ref{Tab:1_acc}, \ref{Tab:3_acc} and \ref{Tab:5_acc}. We see a clear increase in accuracy when the number of layers is increased, especially when we go from a single layer to three layers. The performance of the ansatz with five layers is slightly more than when three layers are used. This is corroborated by Fig. \ref{fig:expressibility-analysis}, where we see a marked increase in the expressibility of our circuit when the number of layers is increased from one to three but not a lot of increase when going from three layers to five. In pairs of classes where one layer of the ansatz performed poorly, like coat vs. shirt or sandals vs. sneakers, we see an appreciable increase in accuracy with an increase in layers. Possibly, this happens because the layered structure allows correlation to be distributed more effectively among the qubits allowing the system to evolve to states that were not previously possible. More explicitly, the ansatz becomes more expressible with each layer, and its overall entangling power also gets enhanced. This can be easily seen in the results of the entangling power analysis as shown in Fig. \ref{fig:entanglingbility-analysis}, where entanglement measures for both TTN and MERA follow a similar exponential trend of improvement with each layer before plateauing down.

In the quantum phase recognition tasks, we first use a VQE-based variational routine with a hardware-efficient ansatz to prepare these systems in the ground state of the Hamiltonian for each spin system instance. This enables us to take care of the sign problem by employing the hybrid quantum-classical routine. Furthermore, the TTN and MERA tensor network ansatz results indicate their effectiveness at solving many-body physics problems. We see that it was much easier for tensor-network-based ans\"{a}tze to classify phases for the transverse-field Ising model, which has simpler interaction terms than the XXZ Heisenberg models. This is in agreement with the previous results obtained for these two models \cite{PhysRevA.102.012415}. Moreover, for the both models, the results for systems on one-dimensional linear lattices were better than the results for two-dimensional rectangular lattices. This is again due to fewer interacting terms, as seen in the previous observation. Among the two tensor-network-based ans\"{a}tze, we find MERA-based ansatz to be overall superior in performance for such tasks than the TTN one, except for the case with the Ising model with four spins. While we can still attribute the MERA-based ansatz's better performance to it being more expressible and generating more entanglement in the states it evolves, the exception tells us that the order in which the correlation gets distributed also matters, especially when the circuit is shallow. In this particular case, it appears from the ans\"{a}tze structures presented in Fig. \ref{fig:4-spin-ttn-mera} that the correlation involving the first and fourth qubits need to be spread before the second and third qubits are entangled. Modifying the MERA ansatz structure by changing the first unitary block to act on the first and fourth qubits as shown in Fig. \ref{fig:4-spin-mera-mod} results in a significant improvement in performance to $99.2 \pm 0.05$ and $99.3 \pm 0.09$ for 1-D and 2-D Ising models, respectively. Finally, we also executed our classifiers on the actual quantum hardware, IBMQ Nairobi, for classifying phases of four spin systems for both 1-D and 2-D cases. We see that even though there's a decreased performance due to the noise present on the device, it was still able to classify the phases decently (Fig. \ref{fig:ising-pred-prob-hardware}). We speculate that this performance can be further improved by employing specific error mitigation techniques like those available in Mitiq \cite{2020arXiv200904417L}. 

Overall, our studies have shown promising results in both tasks, and we conclude that tensor-network-inspired ansatz is an ideal candidate for quantum-enhanced learning of both quantum and classical data. For quantum data, further studies need to be done on tasks such as the phase recognition task on larger, more complicated systems, like systems with 16 or 24 spins, to see how scalable our current model is, which is something we are currently pursuing. On the other hand, for the classical data, more specifically, for the image classification tasks, more work is required to study higher resolution images that would require much better encoding strategies, which is another area of our interest.

\section*{Data Availability}
The code created to run the presented simulations and any related supplementary data could be made available to any reader upon reasonable request.

\vspace{-6pt}
\section*{Acknowledgements}
We acknowledge the use of IBM Quantum services for this work. The views expressed are those of the authors and do not reflect the official policy or position of IBM or the IBM Quantum team.

\bibliographystyle{apsrev4-2}

\bibliography{mlqtn}

\begin{thebibliography}{42}%
\makeatletter
\providecommand \@ifxundefined [1]{%
 \@ifx{#1\undefined}
}%
\providecommand \@ifnum [1]{%
 \ifnum #1\expandafter \@firstoftwo
 \else \expandafter \@secondoftwo
 \fi
}%
\providecommand \@ifx [1]{%
 \ifx #1\expandafter \@firstoftwo
 \else \expandafter \@secondoftwo
 \fi
}%
\providecommand \natexlab [1]{#1}%
\providecommand \enquote  [1]{``#1''}%
\providecommand \bibnamefont  [1]{#1}%
\providecommand \bibfnamefont [1]{#1}%
\providecommand \citenamefont [1]{#1}%
\providecommand \href@noop [0]{\@secondoftwo}%
\providecommand \href [0]{\begingroup \@sanitize@url \@href}%
\providecommand \@href[1]{\@@startlink{#1}\@@href}%
\providecommand \@@href[1]{\endgroup#1\@@endlink}%
\providecommand \@sanitize@url [0]{\catcode `\\12\catcode `\$12\catcode
  `\&12\catcode `\#12\catcode `\^12\catcode `\_12\catcode `\%12\relax}%
\providecommand \@@startlink[1]{}%
\providecommand \@@endlink[0]{}%
\providecommand \url  [0]{\begingroup\@sanitize@url \@url }%
\providecommand \@url [1]{\endgroup\@href {#1}{\urlprefix }}%
\providecommand \urlprefix  [0]{URL }%
\providecommand \Eprint [0]{\href }%
\providecommand \doibase [0]{https://doi.org/}%
\providecommand \selectlanguage [0]{\@gobble}%
\providecommand \bibinfo  [0]{\@secondoftwo}%
\providecommand \bibfield  [0]{\@secondoftwo}%
\providecommand \translation [1]{[#1]}%
\providecommand \BibitemOpen [0]{}%
\providecommand \bibitemStop [0]{}%
\providecommand \bibitemNoStop [0]{.\EOS\space}%
\providecommand \EOS [0]{\spacefactor3000\relax}%
\providecommand \BibitemShut  [1]{\csname bibitem#1\endcsname}%
\let\auto@bib@innerbib\@empty
\bibitem [{\citenamefont {Deng}\ and\ \citenamefont {Li}(2013)}]{Deng2013}%
  \BibitemOpen
  \bibfield  {author} {\bibinfo {author} {\bibfnamefont {L.}~\bibnamefont
  {Deng}}\ and\ \bibinfo {author} {\bibfnamefont {X.}~\bibnamefont {Li}},\
  }\bibfield  {title} {\emph {\bibinfo {title} {{Machine Learning Paradigms for
  Speech Recognition: An Overview}}},\ }\href
  {https://doi.org/10.1109/tasl.2013.2244083} {\bibfield  {journal} {\bibinfo
  {journal} {{IEEE} Transactions on Audio, Speech, and Language Processing}\
  }\textbf {\bibinfo {volume} {21}},\ \bibinfo {pages} {1060} (\bibinfo {year}
  {2013})}\BibitemShut {NoStop}%
\bibitem [{\citenamefont {Popli}\ \emph {et~al.}(2021)\citenamefont {Popli},
  \citenamefont {Tandon}, \citenamefont {Engelsma}, \citenamefont {Onoe},
  \citenamefont {Okubo},\ and\ \citenamefont
  {Namboodiri}}]{https://doi.org/10.48550/arxiv.2104.03255}%
  \BibitemOpen
  \bibfield  {author} {\bibinfo {author} {\bibfnamefont {A.}~\bibnamefont
  {Popli}}, \bibinfo {author} {\bibfnamefont {S.}~\bibnamefont {Tandon}},
  \bibinfo {author} {\bibfnamefont {J.~J.}\ \bibnamefont {Engelsma}}, \bibinfo
  {author} {\bibfnamefont {N.}~\bibnamefont {Onoe}}, \bibinfo {author}
  {\bibfnamefont {A.}~\bibnamefont {Okubo}},\ and\ \bibinfo {author}
  {\bibfnamefont {A.}~\bibnamefont {Namboodiri}},\ }\href
  {https://doi.org/10.48550/ARXIV.2104.03255} {\bibinfo {title} {{A Unified
  Model for Fingerprint Authentication and Presentation Attack Detection}}}
  (\bibinfo {year} {2021})\BibitemShut {NoStop}%
\bibitem [{\citenamefont {Cire\c{s}an}\ \emph {et~al.}(2011)\citenamefont
  {Cire\c{s}an}, \citenamefont {Meier}, \citenamefont {Masci}, \citenamefont
  {Gambardella},\ and\ \citenamefont
  {Schmidhuber}}]{https://doi.org/10.48550/arxiv.1102.0183}%
  \BibitemOpen
  \bibfield  {author} {\bibinfo {author} {\bibfnamefont {D.~C.}\ \bibnamefont
  {Cire\c{s}an}}, \bibinfo {author} {\bibfnamefont {U.}~\bibnamefont {Meier}},
  \bibinfo {author} {\bibfnamefont {J.}~\bibnamefont {Masci}}, \bibinfo
  {author} {\bibfnamefont {L.~M.}\ \bibnamefont {Gambardella}},\ and\ \bibinfo
  {author} {\bibfnamefont {J.}~\bibnamefont {Schmidhuber}},\ }\bibfield
  {title} {\emph {\bibinfo {title} {{High-Performance Neural Networks for
  Visual Object Classification}}},\ }\href@noop {} {\bibfield  {journal}
  {\bibinfo  {journal} {arXiv e-prints}\ } (\bibinfo {year} {2011})},\ \Eprint
  {https://arxiv.org/abs/1102.0183} {arXiv:1102.0183 [cs.AI]} \BibitemShut
  {NoStop}%
\bibitem [{\citenamefont {Amrane}\ \emph {et~al.}(2018)\citenamefont {Amrane},
  \citenamefont {Oukid}, \citenamefont {Gagaoua},\ and\ \citenamefont
  {Ensari}}]{Amrane2018}%
  \BibitemOpen
  \bibfield  {author} {\bibinfo {author} {\bibfnamefont {M.}~\bibnamefont
  {Amrane}}, \bibinfo {author} {\bibfnamefont {S.}~\bibnamefont {Oukid}},
  \bibinfo {author} {\bibfnamefont {I.}~\bibnamefont {Gagaoua}},\ and\ \bibinfo
  {author} {\bibfnamefont {T.}~\bibnamefont {Ensari}},\ }in\ \href
  {https://doi.org/10.1109/ebbt.2018.8391453} {\emph {\bibinfo {booktitle}
  {2018 Electric Electronics, Computer Science, Biomedical
  Engineerings{\textquotesingle} Meeting ({EBBT})}}}\ (\bibinfo  {publisher}
  {{IEEE}},\ \bibinfo {year} {2018})\BibitemShut {NoStop}%
\bibitem [{\citenamefont {Carrasquilla}\ and\ \citenamefont
  {Melko}(2017)}]{Carrasquilla2017}%
  \BibitemOpen
  \bibfield  {author} {\bibinfo {author} {\bibfnamefont {J.}~\bibnamefont
  {Carrasquilla}}\ and\ \bibinfo {author} {\bibfnamefont {R.~G.}\ \bibnamefont
  {Melko}},\ }\bibfield  {title} {\emph {\bibinfo {title} {{Machine learning
  phases of matter}}},\ }\href {https://doi.org/10.1038/nphys4035} {\bibfield
  {journal} {\bibinfo  {journal} {Nature Physics}\ }\textbf {\bibinfo {volume}
  {13}},\ \bibinfo {pages} {431} (\bibinfo {year} {2017})}\BibitemShut
  {NoStop}%
\bibitem [{\citenamefont {van Nieuwenburg}\ \emph {et~al.}(2017)\citenamefont
  {van Nieuwenburg}, \citenamefont {Liu},\ and\ \citenamefont
  {Huber}}]{vanNieuwenburg2017}%
  \BibitemOpen
  \bibfield  {author} {\bibinfo {author} {\bibfnamefont {E.~P.~L.}\
  \bibnamefont {van Nieuwenburg}}, \bibinfo {author} {\bibfnamefont {Y.-H.}\
  \bibnamefont {Liu}},\ and\ \bibinfo {author} {\bibfnamefont {S.~D.}\
  \bibnamefont {Huber}},\ }\bibfield  {title} {\emph {\bibinfo {title}
  {{Learning phase transitions by confusion}}},\ }\href
  {https://doi.org/10.1038/nphys4037} {\bibfield  {journal} {\bibinfo
  {journal} {Nature Physics}\ }\textbf {\bibinfo {volume} {13}},\ \bibinfo
  {pages} {435} (\bibinfo {year} {2017})}\BibitemShut {NoStop}%
\bibitem [{\citenamefont {Loh}\ \emph {et~al.}(1990)\citenamefont {Loh},
  \citenamefont {Gubernatis}, \citenamefont {Scalettar}, \citenamefont {White},
  \citenamefont {Scalapino},\ and\ \citenamefont {Sugar}}]{PhysRevB.41.9301}%
  \BibitemOpen
  \bibfield  {author} {\bibinfo {author} {\bibfnamefont {E.~Y.}\ \bibnamefont
  {Loh}}, \bibinfo {author} {\bibfnamefont {J.~E.}\ \bibnamefont {Gubernatis}},
  \bibinfo {author} {\bibfnamefont {R.~T.}\ \bibnamefont {Scalettar}}, \bibinfo
  {author} {\bibfnamefont {S.~R.}\ \bibnamefont {White}}, \bibinfo {author}
  {\bibfnamefont {D.~J.}\ \bibnamefont {Scalapino}},\ and\ \bibinfo {author}
  {\bibfnamefont {R.~L.}\ \bibnamefont {Sugar}},\ }\bibfield  {title} {\emph
  {\bibinfo {title} {{Sign problem in the numerical simulation of many-electron
  systems}}},\ }\href {https://doi.org/10.1103/PhysRevB.41.9301} {\bibfield
  {journal} {\bibinfo  {journal} {Phys. Rev. B}\ }\textbf {\bibinfo {volume}
  {41}},\ \bibinfo {pages} {9301} (\bibinfo {year} {1990})}\BibitemShut
  {NoStop}%
\bibitem [{\citenamefont {Farhi}\ and\ \citenamefont
  {Neven}(2018)}]{https://doi.org/10.48550/arxiv.1802.06002}%
  \BibitemOpen
  \bibfield  {author} {\bibinfo {author} {\bibfnamefont {E.}~\bibnamefont
  {Farhi}}\ and\ \bibinfo {author} {\bibfnamefont {H.}~\bibnamefont {Neven}},\
  }\href {https://doi.org/10.48550/ARXIV.1802.06002} {\bibinfo {title}
  {{Classification with Quantum Neural Networks on Near Term Processors}}}
  (\bibinfo {year} {2018})\BibitemShut {NoStop}%
\bibitem [{\citenamefont {Peruzzo}\ \emph {et~al.}(2014)\citenamefont
  {Peruzzo}, \citenamefont {McClean}, \citenamefont {Shadbolt}, \citenamefont
  {Yung}, \citenamefont {Zhou}, \citenamefont {Love}, \citenamefont
  {Aspuru-Guzik},\ and\ \citenamefont {O'Brien}}]{Peruzzo2014}%
  \BibitemOpen
  \bibfield  {author} {\bibinfo {author} {\bibfnamefont {A.}~\bibnamefont
  {Peruzzo}}, \bibinfo {author} {\bibfnamefont {J.}~\bibnamefont {McClean}},
  \bibinfo {author} {\bibfnamefont {P.}~\bibnamefont {Shadbolt}}, \bibinfo
  {author} {\bibfnamefont {M.-H.}\ \bibnamefont {Yung}}, \bibinfo {author}
  {\bibfnamefont {X.-Q.}\ \bibnamefont {Zhou}}, \bibinfo {author}
  {\bibfnamefont {P.~J.}\ \bibnamefont {Love}}, \bibinfo {author}
  {\bibfnamefont {A.}~\bibnamefont {Aspuru-Guzik}},\ and\ \bibinfo {author}
  {\bibfnamefont {J.~L.}\ \bibnamefont {O'Brien}},\ }\bibfield  {title} {\emph
  {\bibinfo {title} {{A variational eigenvalue solver on a photonic quantum
  processor}}},\ }\bibfield  {journal} {\bibinfo  {journal} {Nature
  Communications}\ }\textbf {\bibinfo {volume} {5}},\ \href
  {https://doi.org/10.1038/ncomms5213} {10.1038/ncomms5213} (\bibinfo {year}
  {2014})\BibitemShut {NoStop}%
\bibitem [{\citenamefont {Bravo-Prieto}\ \emph {et~al.}(2019)\citenamefont
  {Bravo-Prieto}, \citenamefont {LaRose}, \citenamefont {Cerezo}, \citenamefont
  {Subasi}, \citenamefont {Cincio},\ and\ \citenamefont
  {Coles}}]{2019arXiv190905820B}%
  \BibitemOpen
  \bibfield  {author} {\bibinfo {author} {\bibfnamefont {C.}~\bibnamefont
  {Bravo-Prieto}}, \bibinfo {author} {\bibfnamefont {R.}~\bibnamefont
  {LaRose}}, \bibinfo {author} {\bibfnamefont {M.}~\bibnamefont {Cerezo}},
  \bibinfo {author} {\bibfnamefont {Y.}~\bibnamefont {Subasi}}, \bibinfo
  {author} {\bibfnamefont {L.}~\bibnamefont {Cincio}},\ and\ \bibinfo {author}
  {\bibfnamefont {P.~J.}\ \bibnamefont {Coles}},\ }\bibfield  {title} {\emph
  {\bibinfo {title} {{Variational Quantum Linear Solver}}},\ }\href@noop {}
  {\bibfield  {journal} {\bibinfo  {journal} {arXiv e-prints}\ } (\bibinfo
  {year} {2019})},\ \Eprint {https://arxiv.org/abs/1909.05820}
  {arXiv:1909.05820 [quant-ph]} \BibitemShut {NoStop}%
\bibitem [{\citenamefont {Preskill}(2018)}]{Preskill2018}%
  \BibitemOpen
  \bibfield  {author} {\bibinfo {author} {\bibfnamefont {J.}~\bibnamefont
  {Preskill}},\ }\bibfield  {title} {\emph {\bibinfo {title} {{Quantum
  Computing in the NISQ era and beyond}}},\ }\href
  {https://doi.org/10.22331/q-2018-08-06-79} {\bibfield  {journal} {\bibinfo
  {journal} {Quantum}\ }\textbf {\bibinfo {volume} {2}},\ \bibinfo {pages} {79}
  (\bibinfo {year} {2018})}\BibitemShut {NoStop}%
\bibitem [{\citenamefont {Cerezo}\ \emph {et~al.}(2020)\citenamefont {Cerezo},
  \citenamefont {Arrasmith}, \citenamefont {Babbush}, \citenamefont {Benjamin},
  \citenamefont {Endo}, \citenamefont {Fujii}, \citenamefont {McClean},
  \citenamefont {Mitarai}, \citenamefont {Yuan}, \citenamefont {Cincio},\ and\
  \citenamefont {Coles}}]{2020arXiv201209265C}%
  \BibitemOpen
  \bibfield  {author} {\bibinfo {author} {\bibfnamefont {M.}~\bibnamefont
  {Cerezo}}, \bibinfo {author} {\bibfnamefont {A.}~\bibnamefont {Arrasmith}},
  \bibinfo {author} {\bibfnamefont {R.}~\bibnamefont {Babbush}}, \bibinfo
  {author} {\bibfnamefont {S.~C.}\ \bibnamefont {Benjamin}}, \bibinfo {author}
  {\bibfnamefont {S.}~\bibnamefont {Endo}}, \bibinfo {author} {\bibfnamefont
  {K.}~\bibnamefont {Fujii}}, \bibinfo {author} {\bibfnamefont {J.~R.}\
  \bibnamefont {McClean}}, \bibinfo {author} {\bibfnamefont {K.}~\bibnamefont
  {Mitarai}}, \bibinfo {author} {\bibfnamefont {X.}~\bibnamefont {Yuan}},
  \bibinfo {author} {\bibfnamefont {L.}~\bibnamefont {Cincio}},\ and\ \bibinfo
  {author} {\bibfnamefont {P.~J.}\ \bibnamefont {Coles}},\ }\bibfield  {title}
  {\emph {\bibinfo {title} {{Variational Quantum Algorithms}}},\ }\href@noop {}
  {\bibfield  {journal} {\bibinfo  {journal} {arXiv e-prints}\ } (\bibinfo
  {year} {2020})},\ \Eprint {https://arxiv.org/abs/2012.09265}
  {arXiv:2012.09265 [quant-ph]} \BibitemShut {NoStop}%
\bibitem [{\citenamefont {Azad}\ and\ \citenamefont {Singh}(2022)}]{Azad2022}%
  \BibitemOpen
  \bibfield  {author} {\bibinfo {author} {\bibfnamefont {U.}~\bibnamefont
  {Azad}}\ and\ \bibinfo {author} {\bibfnamefont {H.}~\bibnamefont {Singh}},\
  }\bibfield  {title} {\emph {\bibinfo {title} {{Quantum chemistry calculations
  using energy derivatives on quantum computers}}},\ }\href
  {https://doi.org/10.1016/j.chemphys.2022.111506} {\bibfield  {journal}
  {\bibinfo  {journal} {Chemical Physics}\ }\textbf {\bibinfo {volume} {558}},\
  \bibinfo {pages} {111506} (\bibinfo {year} {2022})}\BibitemShut {NoStop}%
\bibitem [{\citenamefont {Azad}\ and\ \citenamefont
  {Sinha}(2022)}]{2022arXiv220502095A}%
  \BibitemOpen
  \bibfield  {author} {\bibinfo {author} {\bibfnamefont {U.}~\bibnamefont
  {Azad}}\ and\ \bibinfo {author} {\bibfnamefont {A.}~\bibnamefont {Sinha}},\
  }\bibfield  {title} {\emph {\bibinfo {title} {{qLEET: Visualizing Loss
  Landscapes, Expressibility, Entangling Power and Training Trajectories for
  Parameterized Quantum Circuits}}},\ }\href@noop {} {\bibfield  {journal}
  {\bibinfo  {journal} {arXiv e-prints}\ } (\bibinfo {year} {2022})},\ \Eprint
  {https://arxiv.org/abs/2205.02095} {arXiv:2205.02095 [quant-ph]} \BibitemShut
  {NoStop}%
\bibitem [{\citenamefont {Xiao}\ \emph {et~al.}(2017)\citenamefont {Xiao},
  \citenamefont {Rasul},\ and\ \citenamefont
  {Vollgraf}}]{https://doi.org/10.48550/arxiv.1708.07747}%
  \BibitemOpen
  \bibfield  {author} {\bibinfo {author} {\bibfnamefont {H.}~\bibnamefont
  {Xiao}}, \bibinfo {author} {\bibfnamefont {K.}~\bibnamefont {Rasul}},\ and\
  \bibinfo {author} {\bibfnamefont {R.}~\bibnamefont {Vollgraf}},\ }\href
  {https://doi.org/10.48550/ARXIV.1708.07747} {\bibinfo {title}
  {{Fashion-MNIST: a Novel Image Dataset for Benchmarking Machine Learning
  Algorithms}}} (\bibinfo {year} {2017})\BibitemShut {NoStop}%
\bibitem [{\citenamefont {Lazzarin}\ \emph {et~al.}(2022)\citenamefont
  {Lazzarin}, \citenamefont {Galli},\ and\ \citenamefont
  {Prati}}]{Lazzarin2022}%
  \BibitemOpen
  \bibfield  {author} {\bibinfo {author} {\bibfnamefont {M.}~\bibnamefont
  {Lazzarin}}, \bibinfo {author} {\bibfnamefont {D.~E.}\ \bibnamefont
  {Galli}},\ and\ \bibinfo {author} {\bibfnamefont {E.}~\bibnamefont {Prati}},\
  }\bibfield  {title} {\emph {\bibinfo {title} {{Multi-class quantum
  classifiers with tensor network circuits for quantum phase recognition}}},\
  }\href {https://doi.org/10.1016/j.physleta.2022.128056} {\bibfield  {journal}
  {\bibinfo  {journal} {Physics Letters A}\ }\textbf {\bibinfo {volume}
  {434}},\ \bibinfo {pages} {128056} (\bibinfo {year} {2022})}\BibitemShut
  {NoStop}%
\bibitem [{\citenamefont {Stoudenmire}\ and\ \citenamefont
  {Schwab}(2016)}]{https://doi.org/10.48550/arxiv.1605.05775}%
  \BibitemOpen
  \bibfield  {author} {\bibinfo {author} {\bibfnamefont {E.}~\bibnamefont
  {Stoudenmire}}\ and\ \bibinfo {author} {\bibfnamefont {D.~J.}\ \bibnamefont
  {Schwab}},\ }in\ \href
  {https://proceedings.neurips.cc/paper/2016/file/5314b9674c86e3f9d1ba25ef9bb32895-Paper.pdf}
  {\emph {\bibinfo {booktitle} {Advances in Neural Information Processing
  Systems}}},\ Vol.~\bibinfo {volume} {29},\ \bibinfo {editor} {edited by\
  \bibinfo {editor} {\bibfnamefont {D.}~\bibnamefont {Lee}}, \bibinfo {editor}
  {\bibfnamefont {M.}~\bibnamefont {Sugiyama}}, \bibinfo {editor}
  {\bibfnamefont {U.}~\bibnamefont {Luxburg}}, \bibinfo {editor} {\bibfnamefont
  {I.}~\bibnamefont {Guyon}},\ and\ \bibinfo {editor} {\bibfnamefont
  {R.}~\bibnamefont {Garnett}}}\ (\bibinfo  {publisher} {Curran Associates,
  Inc.},\ \bibinfo {year} {2016})\BibitemShut {NoStop}%
\bibitem [{\citenamefont {Or\'{u}s}(2014)}]{Ors2014}%
  \BibitemOpen
  \bibfield  {author} {\bibinfo {author} {\bibfnamefont {R.}~\bibnamefont
  {Or\'{u}s}},\ }\bibfield  {title} {\emph {\bibinfo {title} {{A practical
  introduction to tensor networks: Matrix product states and projected
  entangled pair states}}},\ }\href {https://doi.org/10.1016/j.aop.2014.06.013}
  {\bibfield  {journal} {\bibinfo  {journal} {Annals of Physics}\ }\textbf
  {\bibinfo {volume} {349}},\ \bibinfo {pages} {117} (\bibinfo {year}
  {2014})}\BibitemShut {NoStop}%
\bibitem [{\citenamefont {Shi}\ \emph {et~al.}(2006)\citenamefont {Shi},
  \citenamefont {Duan},\ and\ \citenamefont {Vidal}}]{Shi2006}%
  \BibitemOpen
  \bibfield  {author} {\bibinfo {author} {\bibfnamefont {Y.-Y.}\ \bibnamefont
  {Shi}}, \bibinfo {author} {\bibfnamefont {L.-M.}\ \bibnamefont {Duan}},\ and\
  \bibinfo {author} {\bibfnamefont {G.}~\bibnamefont {Vidal}},\ }\bibfield
  {title} {\emph {\bibinfo {title} {{Classical simulation of quantum many-body
  systems with a tree tensor network}}},\ }\bibfield  {journal} {\bibinfo
  {journal} {Physical Review A}\ }\textbf {\bibinfo {volume} {74}},\ \href
  {https://doi.org/10.1103/physreva.74.022320} {10.1103/physreva.74.022320}
  (\bibinfo {year} {2006})\BibitemShut {NoStop}%
\bibitem [{\citenamefont
  {Vidal}(2008)}]{https://doi.org/10.48550/arxiv.quant-ph/0610099}%
  \BibitemOpen
  \bibfield  {author} {\bibinfo {author} {\bibfnamefont {G.}~\bibnamefont
  {Vidal}},\ }\bibfield  {title} {\emph {\bibinfo {title} {{Class of Quantum
  Many-Body States That Can Be Efficiently Simulated}}},\ }\href
  {https://doi.org/10.1103/PhysRevLett.101.110501} {\bibfield  {journal}
  {\bibinfo  {journal} {Phys. Rev. Lett.}\ }\textbf {\bibinfo {volume} {101}},\
  \bibinfo {pages} {110501} (\bibinfo {year} {2008})}\BibitemShut {NoStop}%
\bibitem [{\citenamefont {Huggins}\ \emph {et~al.}(2019)\citenamefont
  {Huggins}, \citenamefont {Patil}, \citenamefont {Mitchell}, \citenamefont
  {Whaley},\ and\ \citenamefont {Stoudenmire}}]{Huggins2019}%
  \BibitemOpen
  \bibfield  {author} {\bibinfo {author} {\bibfnamefont {W.}~\bibnamefont
  {Huggins}}, \bibinfo {author} {\bibfnamefont {P.}~\bibnamefont {Patil}},
  \bibinfo {author} {\bibfnamefont {B.}~\bibnamefont {Mitchell}}, \bibinfo
  {author} {\bibfnamefont {K.~B.}\ \bibnamefont {Whaley}},\ and\ \bibinfo
  {author} {\bibfnamefont {E.~M.}\ \bibnamefont {Stoudenmire}},\ }\bibfield
  {title} {\emph {\bibinfo {title} {{Towards quantum machine learning with
  tensor networks}}},\ }\href {https://doi.org/10.1088/2058-9565/aaea94}
  {\bibfield  {journal} {\bibinfo  {journal} {Quantum Science and Technology}\
  }\textbf {\bibinfo {volume} {4}},\ \bibinfo {pages} {024001} (\bibinfo {year}
  {2019})}\BibitemShut {NoStop}%
\bibitem [{\citenamefont {Haghshenas}\ \emph {et~al.}(2022)\citenamefont
  {Haghshenas}, \citenamefont {Gray}, \citenamefont {Potter},\ and\
  \citenamefont {Chan}}]{Haghshenas2022}%
  \BibitemOpen
  \bibfield  {author} {\bibinfo {author} {\bibfnamefont {R.}~\bibnamefont
  {Haghshenas}}, \bibinfo {author} {\bibfnamefont {J.}~\bibnamefont {Gray}},
  \bibinfo {author} {\bibfnamefont {A.~C.}\ \bibnamefont {Potter}},\ and\
  \bibinfo {author} {\bibfnamefont {G.~K.-L.}\ \bibnamefont {Chan}},\
  }\bibfield  {title} {\emph {\bibinfo {title} {{Variational Power of Quantum
  Circuit Tensor Networks}}},\ }\bibfield  {journal} {\bibinfo  {journal}
  {Physical Review X}\ }\textbf {\bibinfo {volume} {12}},\ \href
  {https://doi.org/10.1103/physrevx.12.011047} {10.1103/physrevx.12.011047}
  (\bibinfo {year} {2022})\BibitemShut {NoStop}%
\bibitem [{\citenamefont {Wall}\ \emph {et~al.}(2021)\citenamefont {Wall},
  \citenamefont {Abernathy},\ and\ \citenamefont {Quiroz}}]{Wall2021}%
  \BibitemOpen
  \bibfield  {author} {\bibinfo {author} {\bibfnamefont {M.~L.}\ \bibnamefont
  {Wall}}, \bibinfo {author} {\bibfnamefont {M.~R.}\ \bibnamefont
  {Abernathy}},\ and\ \bibinfo {author} {\bibfnamefont {G.}~\bibnamefont
  {Quiroz}},\ }\bibfield  {title} {\emph {\bibinfo {title} {{Generative machine
  learning with tensor networks: Benchmarks on near-term quantum computers}}},\
  }\bibfield  {journal} {\bibinfo  {journal} {Physical Review Research}\
  }\textbf {\bibinfo {volume} {3}},\ \href
  {https://doi.org/10.1103/physrevresearch.3.023010}
  {10.1103/physrevresearch.3.023010} (\bibinfo {year} {2021})\BibitemShut
  {NoStop}%
\bibitem [{\citenamefont {Grant}\ \emph {et~al.}(2018)\citenamefont {Grant},
  \citenamefont {Benedetti}, \citenamefont {Cao}, \citenamefont {Hallam},
  \citenamefont {Lockhart}, \citenamefont {Stojevic}, \citenamefont {Green},\
  and\ \citenamefont {Severini}}]{Grant2018}%
  \BibitemOpen
  \bibfield  {author} {\bibinfo {author} {\bibfnamefont {E.}~\bibnamefont
  {Grant}}, \bibinfo {author} {\bibfnamefont {M.}~\bibnamefont {Benedetti}},
  \bibinfo {author} {\bibfnamefont {S.}~\bibnamefont {Cao}}, \bibinfo {author}
  {\bibfnamefont {A.}~\bibnamefont {Hallam}}, \bibinfo {author} {\bibfnamefont
  {J.}~\bibnamefont {Lockhart}}, \bibinfo {author} {\bibfnamefont
  {V.}~\bibnamefont {Stojevic}}, \bibinfo {author} {\bibfnamefont {A.~G.}\
  \bibnamefont {Green}},\ and\ \bibinfo {author} {\bibfnamefont
  {S.}~\bibnamefont {Severini}},\ }\bibfield  {title} {\emph {\bibinfo {title}
  {{Hierarchical quantum classifiers}}},\ }\bibfield  {journal} {\bibinfo
  {journal} {npj Quantum Information}\ }\textbf {\bibinfo {volume} {4}},\ \href
  {https://doi.org/10.1038/s41534-018-0116-9} {10.1038/s41534-018-0116-9}
  (\bibinfo {year} {2018})\BibitemShut {NoStop}%
\bibitem [{\citenamefont {Yen-Chi~Chen}\ \emph {et~al.}(2020)\citenamefont
  {Yen-Chi~Chen}, \citenamefont {Huang}, \citenamefont {Hsing},\ and\
  \citenamefont {Kao}}]{2020arXiv201114651Y}%
  \BibitemOpen
  \bibfield  {author} {\bibinfo {author} {\bibfnamefont {S.}~\bibnamefont
  {Yen-Chi~Chen}}, \bibinfo {author} {\bibfnamefont {C.-M.}\ \bibnamefont
  {Huang}}, \bibinfo {author} {\bibfnamefont {C.-W.}\ \bibnamefont {Hsing}},\
  and\ \bibinfo {author} {\bibfnamefont {Y.-J.}\ \bibnamefont {Kao}},\
  }\bibfield  {title} {\emph {\bibinfo {title} {{Hybrid quantum-classical
  classifier based on tensor network and variational quantum circuit}}},\
  }\href@noop {} {\bibfield  {journal} {\bibinfo  {journal} {arXiv e-prints}\ }
  (\bibinfo {year} {2020})},\ \Eprint {https://arxiv.org/abs/2011.14651}
  {arXiv:2011.14651} \BibitemShut {NoStop}%
\bibitem [{\citenamefont {Uvarov}\ \emph {et~al.}(2020)\citenamefont {Uvarov},
  \citenamefont {Kardashin},\ and\ \citenamefont
  {Biamonte}}]{PhysRevA.102.012415}%
  \BibitemOpen
  \bibfield  {author} {\bibinfo {author} {\bibfnamefont {A.~V.}\ \bibnamefont
  {Uvarov}}, \bibinfo {author} {\bibfnamefont {A.~S.}\ \bibnamefont
  {Kardashin}},\ and\ \bibinfo {author} {\bibfnamefont {J.~D.}\ \bibnamefont
  {Biamonte}},\ }\bibfield  {title} {\emph {\bibinfo {title} {{Machine learning
  phase transitions with a quantum processor}}},\ }\href
  {https://doi.org/10.1103/PhysRevA.102.012415} {\bibfield  {journal} {\bibinfo
   {journal} {Phys. Rev. A}\ }\textbf {\bibinfo {volume} {102}},\ \bibinfo
  {pages} {012415} (\bibinfo {year} {2020})}\BibitemShut {NoStop}%
\bibitem [{\citenamefont {Parkinson}\ and\ \citenamefont
  {Farnell}(2010)}]{Parkinson2010}%
  \BibitemOpen
  \bibfield  {author} {\bibinfo {author} {\bibfnamefont {J.}~\bibnamefont
  {Parkinson}}\ and\ \bibinfo {author} {\bibfnamefont {D.~J.~J.}\ \bibnamefont
  {Farnell}},\ }\href {https://doi.org/10.1007/978-3-642-13290-2} {\emph
  {\bibinfo {title} {{An Introduction to Quantum Spin Systems}}}}\ (\bibinfo
  {publisher} {Springer Berlin Heidelberg},\ \bibinfo {year}
  {2010})\BibitemShut {NoStop}%
\bibitem [{\citenamefont {Ohanyan}(2015)}]{ohanyan}%
  \BibitemOpen
  \bibfield  {author} {\bibinfo {author} {\bibfnamefont {V.}~\bibnamefont
  {Ohanyan}},\ }\href
  {http://training.hepi.tsu.ge/rtn/activities/sources/Ohanyan.pdf} {\bibinfo
  {title} {{Introduction to quantum spin chains, lecture notes}}} (\bibinfo
  {year} {2015})\BibitemShut {NoStop}%
\bibitem [{\citenamefont {Tan}(2018)}]{atan2018}%
  \BibitemOpen
  \bibfield  {author} {\bibinfo {author} {\bibfnamefont {A.}~\bibnamefont
  {Tan}},\ }\href
  {https://paramekanti.weebly.com/uploads/1/1/2/8/11287579/atan_paper.pdf}
  {\bibinfo {title} {{Quantum Ising Models}}} (\bibinfo {year}
  {2018})\BibitemShut {NoStop}%
\bibitem [{\citenamefont {Bernardi}\ \emph {et~al.}(2022)\citenamefont
  {Bernardi}, \citenamefont {Lazzari},\ and\ \citenamefont
  {Gesmundo}}]{Bernardi2022}%
  \BibitemOpen
  \bibfield  {author} {\bibinfo {author} {\bibfnamefont {A.}~\bibnamefont
  {Bernardi}}, \bibinfo {author} {\bibfnamefont {C.~D.}\ \bibnamefont
  {Lazzari}},\ and\ \bibinfo {author} {\bibfnamefont {F.}~\bibnamefont
  {Gesmundo}},\ }\bibfield  {title} {\emph {\bibinfo {title} {{Dimension of
  tensor network varieties}}},\ }\bibfield  {journal} {\bibinfo  {journal}
  {Communications in Contemporary Mathematics}\ }\href
  {https://doi.org/10.1142/s0219199722500596} {10.1142/s0219199722500596}
  (\bibinfo {year} {2022})\BibitemShut {NoStop}%
\bibitem [{\citenamefont {Fuglede}\ and\ \citenamefont
  {Topsoe}(2004)}]{Fuglede2004}%
  \BibitemOpen
  \bibfield  {author} {\bibinfo {author} {\bibfnamefont {B.}~\bibnamefont
  {Fuglede}}\ and\ \bibinfo {author} {\bibfnamefont {F.}~\bibnamefont
  {Topsoe}},\ }in\ \href {https://doi.org/10.1109/isit.2004.1365067} {\emph
  {\bibinfo {booktitle} {International Symposium on Information Theory, 2004.
  {ISIT} 2004. Proceedings.}}}\ (\bibinfo  {publisher} {{IEEE}},\ \bibinfo
  {year} {2004})\BibitemShut {NoStop}%
\bibitem [{\citenamefont {Jozsa}(1994)}]{Jozsa1994}%
  \BibitemOpen
  \bibfield  {author} {\bibinfo {author} {\bibfnamefont {R.}~\bibnamefont
  {Jozsa}},\ }\bibfield  {title} {\emph {\bibinfo {title} {{Fidelity for Mixed
  Quantum States}}},\ }\href {https://doi.org/10.1080/09500349414552171}
  {\bibfield  {journal} {\bibinfo  {journal} {Journal of Modern Optics}\
  }\textbf {\bibinfo {volume} {41}},\ \bibinfo {pages} {2315} (\bibinfo {year}
  {1994})}\BibitemShut {NoStop}%
\bibitem [{\citenamefont {Meyer}\ and\ \citenamefont
  {Wallach}(2002)}]{Meyer2002}%
  \BibitemOpen
  \bibfield  {author} {\bibinfo {author} {\bibfnamefont {D.~A.}\ \bibnamefont
  {Meyer}}\ and\ \bibinfo {author} {\bibfnamefont {N.~R.}\ \bibnamefont
  {Wallach}},\ }\bibfield  {title} {\emph {\bibinfo {title} {{Global
  entanglement in multiparticle systems}}},\ }\href
  {https://doi.org/10.1063/1.1497700} {\bibfield  {journal} {\bibinfo
  {journal} {Journal of Mathematical Physics}\ }\textbf {\bibinfo {volume}
  {43}},\ \bibinfo {pages} {4273} (\bibinfo {year} {2002})}\BibitemShut
  {NoStop}%
\bibitem [{\citenamefont {LeCun}\ and\ \citenamefont
  {Cortes}(2012)}]{lecun-mnisthandwrittendigit-2010}%
  \BibitemOpen
  \bibfield  {author} {\bibinfo {author} {\bibfnamefont {Y.}~\bibnamefont
  {LeCun}}\ and\ \bibinfo {author} {\bibfnamefont {C.}~\bibnamefont {Cortes}},\
  }\href {http://yann.lecun.com/exdb/mnist/} {\bibinfo {title} {{MNIST
  handwritten digit database}}} (\bibinfo {year} {2012})\BibitemShut {NoStop}%
\bibitem [{\citenamefont {SE}(2017)}]{benchmark_dashboard}%
  \BibitemOpen
  \bibfield  {author} {\bibinfo {author} {\bibfnamefont {Z.}~\bibnamefont
  {SE}},\ }\href {http://fashion-mnist.s3-website.eu-central-1.amazonaws.com/}
  {\bibinfo {title} {{Fashion MNIST Benchmark}}} (\bibinfo {year}
  {2017})\BibitemShut {NoStop}%
\bibitem [{IBM(2021)}]{IBMQ}%
  \BibitemOpen
  \href@noop {} {\bibinfo {title} {{IBM Quantum}}},\ \bibinfo {howpublished}
  {\url{https://quantum-computing.ibm.com/}} (\bibinfo {year}
  {2021})\BibitemShut {NoStop}%
\bibitem [{\citenamefont {Kingma}\ and\ \citenamefont
  {Ba}(2014)}]{https://doi.org/10.48550/arxiv.1412.6980}%
  \BibitemOpen
  \bibfield  {author} {\bibinfo {author} {\bibfnamefont {D.~P.}\ \bibnamefont
  {Kingma}}\ and\ \bibinfo {author} {\bibfnamefont {J.}~\bibnamefont {Ba}},\
  }\bibfield  {title} {\emph {\bibinfo {title} {{Adam: A Method for Stochastic
  Optimization}}},\ }\href
  {https://doi.org/https://hdl.handle.net/11245/1.505367} {\bibfield  {journal}
  {\bibinfo  {journal} {arXiv e-prints}\ ,\ \bibinfo {eid} {arXiv:1412.6980}}
  (\bibinfo {year} {2014})},\ \Eprint {https://arxiv.org/abs/1412.6980}
  {arXiv:1412.6980 [cs.LG]} \BibitemShut {NoStop}%
\bibitem [{\citenamefont {Zhang}\ and\ \citenamefont
  {Sabuncu}(2018)}]{10.5555/3327546.3327555}%
  \BibitemOpen
  \bibfield  {author} {\bibinfo {author} {\bibfnamefont {Z.}~\bibnamefont
  {Zhang}}\ and\ \bibinfo {author} {\bibfnamefont {M.~R.}\ \bibnamefont
  {Sabuncu}},\ }in\ \href {https://doi.org/10.5555/3327546.3327555} {\emph
  {\bibinfo {booktitle} {Proceedings of the 32nd International Conference on
  Neural Information Processing Systems}}},\ \bibinfo {series and number}
  {NIPS'18}\ (\bibinfo  {publisher} {Curran Associates Inc.},\ \bibinfo
  {address} {Red Hook, NY, USA},\ \bibinfo {year} {2018})\ pp.\ \bibinfo
  {pages} {8792--8802}\BibitemShut {NoStop}%
\bibitem [{\citenamefont {Hashizume}\ \emph {et~al.}(2022)\citenamefont
  {Hashizume}, \citenamefont {McCulloch},\ and\ \citenamefont
  {Halimeh}}]{Hashizume2022}%
  \BibitemOpen
  \bibfield  {author} {\bibinfo {author} {\bibfnamefont {T.}~\bibnamefont
  {Hashizume}}, \bibinfo {author} {\bibfnamefont {I.~P.}\ \bibnamefont
  {McCulloch}},\ and\ \bibinfo {author} {\bibfnamefont {J.~C.}\ \bibnamefont
  {Halimeh}},\ }\bibfield  {title} {\emph {\bibinfo {title} {{Dynamical phase
  transitions in the two-dimensional transverse-field Ising model}}},\
  }\bibfield  {journal} {\bibinfo  {journal} {Physical Review Research}\
  }\textbf {\bibinfo {volume} {4}},\ \href
  {https://doi.org/10.1103/physrevresearch.4.013250}
  {10.1103/physrevresearch.4.013250} (\bibinfo {year} {2022})\BibitemShut
  {NoStop}%
\bibitem [{\citenamefont {Franchini}(2017)}]{Franchini2017}%
  \BibitemOpen
  \bibfield  {author} {\bibinfo {author} {\bibfnamefont {F.}~\bibnamefont
  {Franchini}},\ }\href {https://doi.org/10.1007/978-3-319-48487-7} {\emph
  {\bibinfo {title} {{An Introduction to Integrable Techniques for
  One-Dimensional Quantum Systems}}}}\ (\bibinfo  {publisher} {Springer
  International Publishing},\ \bibinfo {year} {2017})\BibitemShut {NoStop}%
\bibitem [{\citenamefont {yer}(2019)}]{Saryer2019}%
  \BibitemOpen
  \bibfield  {author} {\bibinfo {author} {\bibfnamefont {O.~S.~S.}\
  \bibnamefont {yer}},\ }\bibfield  {title} {\emph {\bibinfo {title}
  {{Two-dimensional quantum-spin-1/2 XXZ magnet in zero magnetic field: Global
  thermodynamics from renormalisation group theory}}},\ }\href
  {https://doi.org/10.1080/14786435.2019.1605212} {\bibfield  {journal}
  {\bibinfo  {journal} {Philosophical Magazine}\ }\textbf {\bibinfo {volume}
  {99}},\ \bibinfo {pages} {1787} (\bibinfo {year} {2019})}\BibitemShut
  {NoStop}%
\bibitem [{\citenamefont {LaRose}\ \emph {et~al.}(2020)\citenamefont {LaRose},
  \citenamefont {Mari}, \citenamefont {Kaiser}, \citenamefont {Karalekas},
  \citenamefont {Alves}, \citenamefont {Czarnik},\ and\ \citenamefont
  {et~al.}}]{2020arXiv200904417L}%
  \BibitemOpen
  \bibfield  {author} {\bibinfo {author} {\bibfnamefont {R.}~\bibnamefont
  {LaRose}}, \bibinfo {author} {\bibfnamefont {A.}~\bibnamefont {Mari}},
  \bibinfo {author} {\bibfnamefont {S.}~\bibnamefont {Kaiser}}, \bibinfo
  {author} {\bibfnamefont {P.~J.}\ \bibnamefont {Karalekas}}, \bibinfo {author}
  {\bibfnamefont {A.~A.}\ \bibnamefont {Alves}}, \bibinfo {author}
  {\bibfnamefont {P.}~\bibnamefont {Czarnik}},\ and\ \bibinfo {author}
  {\bibnamefont {et~al.}},\ }\bibfield  {title} {\emph {\bibinfo {title}
  {{Mitiq: A software package for error mitigation on noisy quantum
  computers}}},\ }\href@noop {} {\bibfield  {journal} {\bibinfo  {journal}
  {arXiv e-prints}\ ,\ \bibinfo {eid} {arXiv:2009.04417}} (\bibinfo {year}
  {2020})},\ \Eprint {https://arxiv.org/abs/2009.04417} {arXiv:2009.04417
  [quant-ph]} \BibitemShut {NoStop}%
\end{thebibliography}%


%

\end{document}